\documentclass[twocolumn,final]{svjour3}

\usepackage{graphicx}
\usepackage{color}
\usepackage{amsmath}
\usepackage{mathptmx}
\usepackage{upgreek}
\usepackage{ae}
\usepackage{subfigure}
\usepackage{multirow}
\usepackage[utf8]{inputenc}
\usepackage{natbib}
\usepackage{hyperref}

\newcommand{\micron}{\upmu\text{m}}
\newcommand{\Ht}{\ensuremath{Ht}}

\newcommand{\Htt}{\ensuremath{Ht_\text{t}}}
\newcommand{\Canum}{\ensuremath{Ca}}
\newcommand{\figref}[1]{{Fig.~\ref{#1}}}
\newcommand{\tabref}[1]{{Tab.~\ref{#1}}}
\newcommand{\CFL}{\ell_\text{CFL}}

\newcommand{\tad}{t_\text{ad}}
\newcommand{\WSR}{\dot{\gamma}_\text{w}}
\newcommand{\rRBC}{r_\text{RBC}}

\begin{document}

\sloppy

\title{Effect of tube diameter and capillary number on platelet margination and near-wall dynamics}
% \subtitle{}
%\titlerunning{Short form of title}
\author{Timm Krüger}

\institute{Timm Krüger \at
              University of Edinburgh \\
              School of Engineering \\
              King's Buildings \\
              Mayfield Road \\
              Edinburgh EH9 3FB \\
              Tel.: +44 131-650-5679 \\
              Fax: +44 131-650-6554 \\
              \email{timm.krueger@ed.ac.uk}
}

\date{}

\maketitle

\begin{abstract}
 The effect of tube diameter $D$ and capillary number $\Canum$ on platelet margination in blood flow at $\approx 37\%$ tube haematocrit is investigated.
 The system is modelled as three-dimensional suspension of deformable red blood cells and nearly rigid platelets using a combination of the lattice-Boltzmann, immersed boundary and finite element methods.
 Results show that margination is facilitated by a non-diffusive radial platelet transport.
 This effect is important near the edge of the cell-free layer, but it is only observed for $\Canum > 0.2$, when red blood cells are tank-treading rather than tumbling.
 It is also shown that platelet trapping in the cell-free layer is reversible for $\Canum \leq 0.2$.
 Only for the smallest investigated tube ($D = 10\,\micron$) margination is essentially independent of $\Canum$.
 Once platelets have reached the cell-free layer, they tend to slide rather than tumble.
 The tumbling rate is essentially independent of $\Canum$ but increases with $D$.
 Tumbling is suppressed by the strong confinement due to the relatively small cell-free layer thickness at $\approx 37\%$ tube haematocrit.
%  \subclass{76Z05}
\end{abstract}

\section{Introduction}
\label{intro}

The aim of this article is to investigate the effect of tube diameter and capillary number (wall shear rate) on the margination and near-wall dynamics of platelets in blood flow via highly resolved three-dimensional computer simulations.
Platelet adhesion or activation are not in the focus of this research.

Human blood is, by volume, composed of about $55\%$ plasma (mostly water and proteins) and $45\%$ suspended blood cells.
The majority of the blood cells are red blood cells (RBCs), also called erythrocytes.
Other, less common blood cells are white blood cells (leukocytes, one for 1000 RBCs) which form a significant part of the immune system \citep{robertson_hemorheology_2007} and platelets (thrombocytes, one for 15 RBCs) \citep{almomani_micro-scale_2008}.
The volume fraction of the cellular phase is called haematocrit $\Ht$.

Platelets play an important role in the process of blood clotting and the repair of damaged vessel walls (endothelium).
By mechanical contact with damaged endothelium, platelets can be activated.
The activation involves a reorganisation of the platelet cytoskeleton, making them more flexible \citep{fogelson_immersed-boundary-type_2008}.
Activated platelets are capable of forming networks via fibrinogen and von Willebrand factor molecules giving rise to the build-up of blood clots \citep{kulkarni_revised_2000, doggett_selectin-like_2002}.
Non-activated platelets can be considered as rigid discoid particles over the entire physiological shear stress range (up to a few $10\, \text{Pa}$) \citep{goldsmith_microrheology_1967, teirlinck_orientation_1984, turitto_rheology_1992}.

The main task of platelets is to recognise regions of damaged endothelium and to assist in its repair.
Indeed, it is observed that the platelet concentration is, under certain conditions, significantly increased close to the walls \citep{turitto_platelet_1972, goldsmith_rheological_1986}.
This near-wall excess is of physiological importance as it increases the probability for the platelets to adhere to the damaged endothelium \citep{yeh_transient_1994}.
Platelet concentration near the wall can be several times higher than that at the centre of the flow \citep{tangelder_localization_1982, tilles_near-wall_1987, aarts_blood_1988, eckstein_concentration_1989}.
The lateral motion of platelets, leading to this concentration inhomogeneity, is called \emph{margination}.
Usually, the near-wall concentration peak is located within or near the cell-free layer (CFL) which is depleted of RBCs \citep{yeh_transient_1994}.
The CFL thickness $\CFL$ plays an important role in blood flow, leading to the Fåhræus and Fåhræus-Lindqvist effects \citep{faahraeus_viscosity_1931, lei_blood_2013}.

RBCs have a twofold effect on the platelet motion.
First, it has been observed that the presence of RBCs increases the diffusivity of platelets in blood vessels by orders of magnitude compared to their Brownian diffusivity \citep{aarts_red_1983, aarts_red_1984, breugel_role_1992, turitto_rheology_1992}.
Furthermore, non-diffusive platelet margination towards the vessel wall increases strongly when RBCs are present \citep{turitto_red_1980, cadroy_effects_1990, joist_platelet_1998, peerschke_ex_2004}.

The wall shear stress and the haematocrit have been identified as important parameters for margination \citep{turitto_platelet_1975, turitto_platelet_1979, turitto_rheology_1992, zhao_investigation_2007, fogelson_immersed-boundary-type_2008, reasor_determination_2013}.
It is known that margination is only significant when the wall shear rate is above $200\,\text{s}^{-1}$ \citep{tilles_near-wall_1987, eckstein_conditions_1988, bilsker_freeze-capture_1989}.
There also seems to be a minimum haematocrit of $10\%$ below which no margination is observed for physiological stresses \citep{tilles_near-wall_1987, waters_concentration_1990}.
Margination becomes more important with increasing haematocrit \citep{yeh_transient_1994}.
Also platelet size and shape play a role for the margination efficiency \citep{eckstein_conditions_1988, thompson_margination_2013, reasor_determination_2013, muller_margination_2014}.

Only in recent years, resolved computer simulations became available to study the mechanisms of platelet margination directly, without assuming effective platelet transport models.
Simulations are able to provide microscopic details (such as stresses, cell velocities and deformations) that are difficult to measure in experiments.
As already assumed in the 1970s by \cite{goldsmith_red_1971} and \cite{turitto_platelet_1972}, it is nowadays generally accepted that the dynamics of RBCs and their hydrodynamic interactions with platelets under the influence of shear are responsible for margination, rather than a volume exclusion effect \citep{crowl_computational_2010}.
\cite{zhao_shear-induced_2011} found that the lateral platelet motion in the RBC-rich region is diffusive and caused by shear-induced fluid velocity fluctuations.
According to \cite{kumar_mechanism_2012}, platelet margination, at least in the dilute regime, is caused by heterogeneous collisions of platelets and RBCs.
If the RBCs are sufficiently soft, these collisions lead to directed events which are necessary to explain margination \citep{eckstein_model_1991}.
It can be concluded that the dynamical state of RBCs (tumbling versus tank-treading) plays a major role in platelet margination, as already suggested by \cite{yeh_transient_1994}.
Recent review articles provide more details \citep{kumar_margination_2012, fogelson_fluid_2015}.

In general, platelet margination is still not well understood.
This research aims at investigating in more detail the effect of tube diameter and capillary number on margination.
Three-dimensional computer simulations employing the lattice-Boltzmann, immersed boundary and finite element methods (section \ref{sec:numerics}) are performed to investigate platelet trajectories in an environment of deformable RBCs.
The relevant parameter space and the system setup are presented in section \ref{sec:simulations}.
The results about platelet margination and platelet dynamics in the CFL are presented and discussed in section \ref{sec:results}.
Finally, the observations and implications are summarised in section \ref{sec:conclusions}.

\section{Physical model and numerical methods}
\label{sec:numerics}

The numerical methods employed in this work are briefly outlined.
Details and benchmark tests are reported elsewhere \citep{timm_kruger_computer_2011, kruger_efficient_2011, kruger_crossover_2013, kruger_deformability-based_2014}.
The lattice-Boltzmann method (section \ref{sec:LBM}) is used as Navier-Stokes solver, and the immersed boundary method (section \ref{sec:IBM}) assumes the role of the fluid-structure interaction.
For the membranes of the red blood cells, a finite element approach (section \ref{sec:membrane}) is employed.
In section \ref{sec:viscosity}, the numerical implementation of the viscosity contrast is outlined.
Tab.~\ref{tab:symbols} provides an overview of all relevant symbols and parameters.

\subsection{Lattice-Boltzmann method}
\label{sec:LBM}

The standard lattice-Boltzmann method with D3Q19 lattice \citep{qian_lattice_1992} and BGK collision operator \citep{bhatnagar_model_1954} is employed to solve the Navier-Stokes equations \citep{succi_lattice_2001, aidun_lattice-boltzmann_2010}.

The BGK operator is parametrised by a relaxation time $\tau \Delta t$ (with $\Delta t$ being the time step).
The kinematic fluid viscosity is directly related to $\tau$ according to
\begin{equation}
 \label{eq:lbm_viscosity}
 \nu = \frac{1}{3} \left(\tau - \frac{1}{2}\right) \frac{\Delta x^2}{\Delta t}
\end{equation}
where $\Delta x$ is the lattice resolution.

External forces, such as those from the immersed boundary method (section~\ref{sec:IBM}), are coupled through the forcing scheme by Shan and Chen \citep{shan_lattice_1993}.

\subsection{Immersed boundary method}
\label{sec:IBM}

The immersed boundary method (IBM) \citep{peskin_flow_1972, peskin_immersed_2002} is used to couple the fluid flow on the Eulerian lattice and the off-lattice membrane interface dynamics.
IBM has often been employed for the simulation of elastic deformable objects, e.g.~by \cite{eggleton_large_1998, zhang_immersed_2007, sui_hybrid_2008, fogelson_immersed-boundary-type_2008, doddi_three-dimensional_2009}.

Fluid velocities are interpolated at the locations of the membrane vertices, and membrane forces (section~\ref{sec:membrane}) are distributed back to the lattice.
For this purpose, a short-range tri-linear interpolation stencil is used that requires only $2^3$ lattice points in three dimensions \citep{peskin_immersed_2002}.

\subsection{Red blood cell mesh and elasticity model}
\label{sec:membrane}

\begin{figure}[t]
 \subfigure[\label{fig:meshes_a}]{\includegraphics[width=0.32\linewidth]{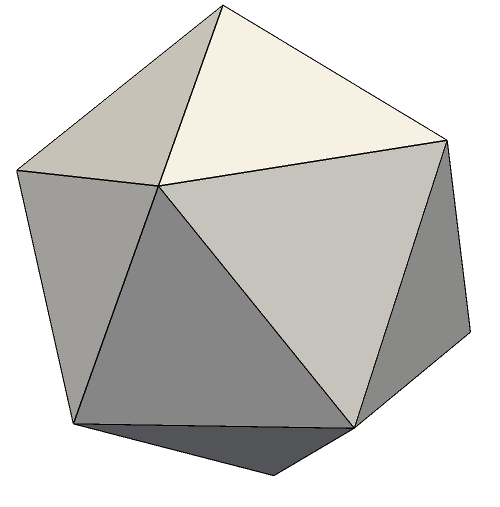}} \hfill
 \subfigure[\label{fig:meshes_b}]{\includegraphics[width=0.32\linewidth]{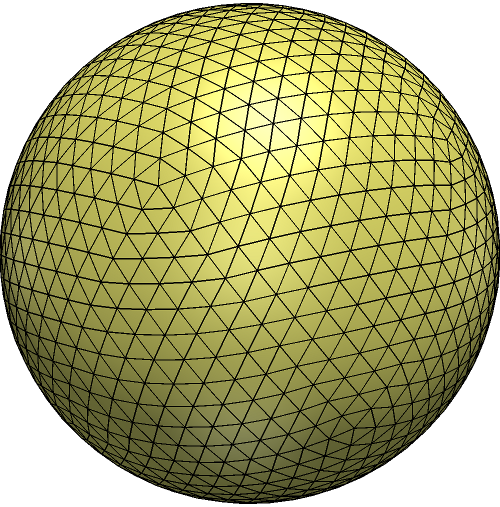}} \hfill
 \subfigure[\label{fig:meshes_c}]{\includegraphics[width=0.32\linewidth]{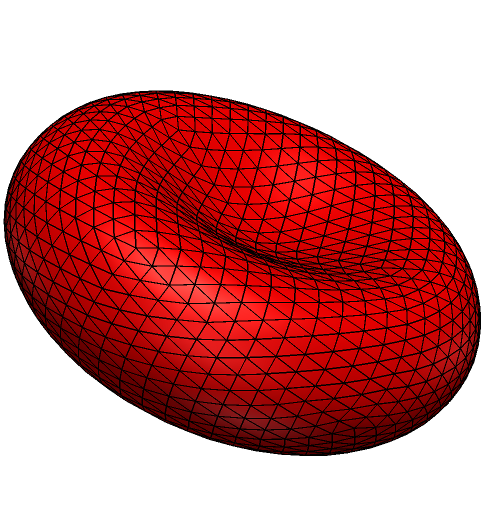}} \\
 \caption{Mesh generation for a red blood cell. The 20 face elements of an icosahedron (a) are subdivided, and the resulting new vertices shifted to the circumsphere (b). The mesh of a red blood cell (c) can be obtained by adjusting the vertex positions in space. The meshes in (b) and (c) have 2880 elements each.}
 \label{fig:meshes}
\end{figure}

In their undeformed state, RBCs are biconcave discs with a diameter of about $8\,\micron$.
RBC membranes are characterised by two major elastic contributions: in-plane shear and bending resistance \citep{skalak_strain_1973, evans_bending_1974}.
Furthermore, local and global area changes are small (typically below 1\%) since the lipid bilayer is incompressible.
The total RBC surface $A^{(0)} \approx 135\, \micron^2$ is therefore essentially constant.
Under physiological conditions, the RBC volume, $V^{(0)} \approx 94\, \micron^3$, is also constant \citep{evans_improved_1972}.
The strong deformability of RBCs is facilitated by the large area excess (additional area compared to a sphere with the same volume) which is about $35\%$.
This way, RBCs can squeeze through capillaries with diameters as small as $4\, \micron$ without violating the surface and volume constraints \citep{skalak_deformation_1969}.

In the present work, RBCs and platelets are treated as closed membranes discretised by $N_\text{f}$ flat faces (area elements).
Each element consists of three nodes (vertices) of which there are in total $N_\text{n} = (N_\text{f} + 4) / 2$.
The generation of the RBC mesh, depicted in Fig.~\ref{fig:meshes}, is based on the icosahedron-refinement procedure \citep{ramanujan_deformation_1998, timm_kruger_computer_2011}.
In the present work, each RBC mesh consists of 2880 elements and 1442 vertices.
Platelets are assumed to be non-activated and modelled as nearly rigid, oblate ellipsoids with an aspect ratio $0.28$ and major radius of $1.8\,\micron$.
Each platelet mesh comprises 320 faces and 162 vertices.

The RBC membrane is modelled as a hyperelastic continuum with four elastic energy contributions as outlined below.
Membrane viscosity is neglected.
It is assumed that undeformed RBCs are stress-free.
The RBC equilibrium shape is imposed through the input geometry in \figref{fig:meshes_c}.

Skalak's constitutive law for the in-plane energy density is employed in the present simulations \citep{skalak_strain_1973}:
\begin{equation}
 \label{eq:skalak}
 E_\text{S} = \int \mathrm{d}A\, \left(\frac{\kappa_\text{S}}{12} \left(I_1^2 + 2 I_1 - 2 I_2\right) + \frac{\kappa_\alpha}{12} I_2^2\right).
\end{equation}
The two parameters $\kappa_\text{S}$ and $\kappa_\alpha$ control the strength of the membrane response to local shear deformation and dilation, respectively.
For healthy RBCs, the values are $\kappa_\text{S} = 5.3\, \upmu\text{N}\, \text{m}^{-1}$ and $\kappa_\alpha = 0.5\, \text{N}\, \text{m}^{-1}$ \citep{gompper_soft_2008}.
The in-plane deformation is described by the two strain invariants: $I_1$ and $I_2$.
Eq.~\eqref{eq:skalak} is discretised on the membrane mesh according to \cite{charrier_free_1989} and \cite{shrivastava_large_1993}.
More details are provided by \cite{kruger_efficient_2011}.

For the bending energy, the discretised form
\begin{equation}
 \label{eq:bending}
 E_{\text{B}} = \frac{\sqrt{3} \kappa_{\text{B}}}{2} \sum_{\left\langle i,j \right\rangle} \left(\theta_{ij} - \theta_{ij}^{(0)}\right)^2
\end{equation}
is used.
Here, $\kappa_\text{B}$ is the bending resistance ($2 \cdot 10^{-19}\, \text{N}\, \text{m}$ for an RBC), $\theta_{ij}$ is the angle between the normal vectors $\hat{\vec{n}}_i$ and $\hat{\vec{n}}_j$ of two neighbouring elements, and $\theta_{ij}^{(0)}$ is the corresponding equilibrium angle of the undeformed membrane.
The sum runs over all pairs of neighbouring elements.

In order to maintain a nearly constant RBC surface area and following \cite{evans_mechanics_1980} and \cite{seifert_configurations_1997}, an additional energy is introduced:
\begin{equation}
 \label{eq:surface}
 E_{\text{A}} = \frac{\kappa_\text{A}}{2} \frac{\left(A - A^{(0)}\right)^2}{A^{(0)}}.
\end{equation}
The magnitude of the surface energy is controlled by the surface modulus $\kappa_\text{A}$.
A similar approach is used for the RBC volume \citep{seifert_configurations_1997}:
\begin{equation}
 \label{eq:volume}
 E_{\text{V}} = \frac{\kappa_\text{V}}{2} \frac{\left(V - V^{(0)}\right)^2}{V^{(0)}}.
\end{equation}

The total membrane energy $E$ is the sum of all contributions $E_\text{S}$, $E_\text{B}$, $E_\text{A}$ and $E_\text{V}$.
The forces acting on membrane vertices at position $\vec{r}_j$ are computed from the energy functional $E(\{\vec{r}_j\})$ via the principle of virtual work:
\begin{equation}
 \label{eq:membraneforce}
 \vec{F}_j = - \frac{\partial E(\{\vec{r}_j\})}{\partial \vec{r}_j}.
\end{equation}
The differentiation is performed analytically, and the resulting forces are implemented in the code.
A detailed derivation of all force contributions is given by \cite{timm_kruger_computer_2011}.

\subsection{Viscosity contrast}
\label{sec:viscosity}

\begin{figure}[bt]
 \includegraphics[width=\linewidth]{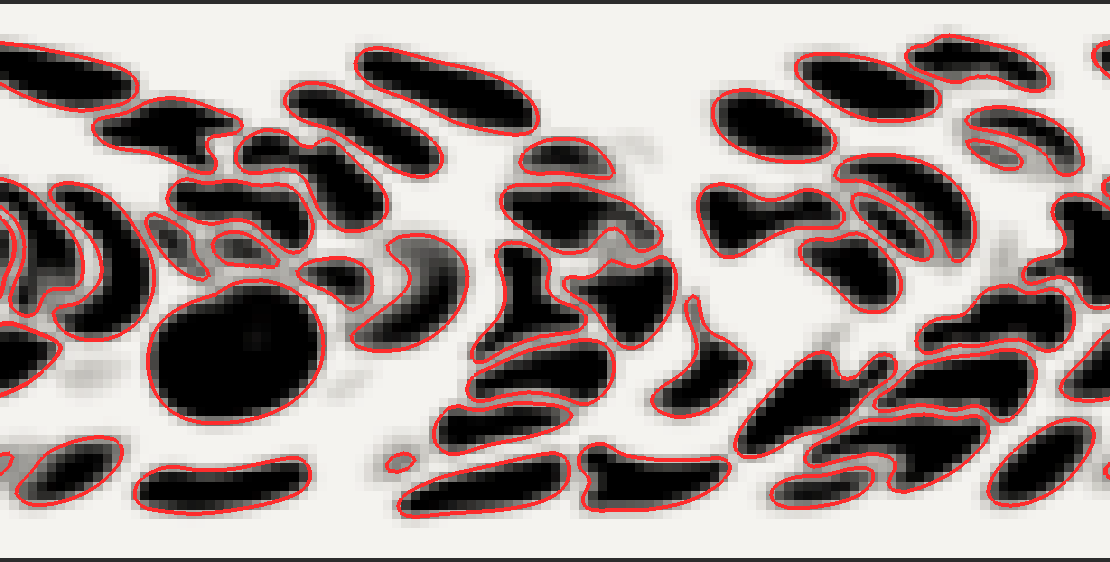}
 \caption{Indicator field $I(\vec{x})$ for distinguishing between interior and exterior fluid sites. A two-dimensional cross-section of a three-dimensional RBC suspension in a tube is shown. Red lines denote the Lagrangian membranes. The grey-scale lattice-like portion of the figure reflects $I(\vec{x})$ on the same cross-section. Black denotes the interior, white the exterior.}
 \label{fig:index_field}
\end{figure}

In order to implement a viscosity contrast $\Lambda$, i.e.~different viscosities inside and outside of the RBCs, each lattice site requires up-to-date information about its location relative to nearby Lagrangian vertices.
A signed distance field as detailed by \cite{frijters_parallelised_2015} is used to construct an indicator field $I(\vec{x})$ at each lattice site $\vec{x}$, denoting the relative location of each site with respect to nearby RBCs and platelets ($I = 0$ outside and $I = 1$ inside particles with a linear slope across the particle surface).
The instantaneous kinematic viscosity is then obtained from the indicator field
\begin{equation}
 \label{eq:viscosity_field}
 \nu(\vec x) = \nu_{\text{out}} \left(1 - I(\vec x)\right) + \nu_{\text{in}} I(\vec x)
\end{equation}
where $\nu_{\text{out}}$ and $\nu_{\text{in}}$ are the external and internal viscosities, respectively, and
\begin{equation}
 \Lambda = \frac{\nu_{\text{in}}}{\nu_{\text{out}}}
\end{equation}
is the viscosity contrast.
Fig.~\ref{fig:index_field} shows the indicator field for a given RBC configuration.

\section{System parameters and simulation setup}
\label{sec:simulations}

The system parameters and the simulation setup are discussed.
After introducing the relevant system parameters in section \ref{sec:physicalparameters}, the choice of simulation parameters is briefly discussed in section \ref{sec:simulationparameters}.
The initialisation of the simulations (section \ref{sec:initialization}) deserves some additional attention.

\subsection{Physical parameters}
\label{sec:physicalparameters}

\begin{table}[b]
 \caption{Overview of used parameters and symbols.}
 \label{tab:symbols}
 \centering
 \begin{tabular}{lll}
  \hline\noalign{\smallskip}
  Parameter & symbol & value \\
  \noalign{\smallskip}\hline\noalign{\smallskip}
  Tube radius, diameter & $R$, $D$ & variable \\
  Tube segment length & $L$ & $48\,\micron$ \\
  RBC radius & $\rRBC$ & $4\,\micron$ \\
  Platelet radius & $r_\text{pl}$ & $1.8\,\micron$ \\
  Platelet thickness & $h_\text{pl}$ & $1\,\micron$ \\
  Tube haematocrit & $\Htt$ & $\approx 37\%$ \\
  RBC count & $N_\text{RBC}$ & variable \\
  Platelet count & $N_\text{pl}$ & $N_\text{pl} = N_\text{RBC} / 2$ \\
  Fluid density & $\rho$ & $1000\,\text{kg}\,\text{m}^{-3}$ \\
  Cytoplasm viscosity & $\eta_\text{in}$ & $5\, \text{mPa}\, \text{s}$ \\
  Plasma viscosity & $\eta_\text{ex}$ & $1\, \text{mPa}\, \text{s}$ \\
  viscosity contrast & $\Lambda$ & $5$ \\
  RBC shear elasticity & $\kappa_\text{S}$ & $5.3\,\upmu\text{N}\,\text{m}^{-1}$ \\
  RBC bending rigidity & $\kappa_\text{B}$ & $2 \cdot 10^{-19}\,\text{N}\,\text{m}$ \\
  Reduced bending modulus & $\kappa_\text{B} / (\kappa_\text{S} r^2)$ & $1 / 424$ \\
  Pressure gradient & $p'$ & variable \\
  Capillary number & $\Canum$ & variable \\
  Lattice resolution & $\Delta x$ & $0.33\,\micron$ \\
  Time step & $\Delta t$ & variable \\
  Centre velocity (no cells) & $\hat{u}_0$ & $0.05\, \Delta x / \Delta t$ \\
  Average velocity (no cells) & $\bar{u}_0$ & $\hat{u}_0 / 2$ \\
  Wall shear rate & $\WSR$ & $4 \hat{u}_0 / D$ \\
  Number of time steps & $N_t$ & $2 \cdot 10^5$ \\
  Advection time & $\tad$ & $960\, \Delta t$ \\
  \noalign{\smallskip}\hline
 \end{tabular}
\end{table}

The present research concerns with blood flow in straight tubes, which involves a series of parameters and related symbolds as collected in \tabref{tab:symbols}.

The flow is driven along the tube axis in positive $x$-direction by a pressure gradient $p'$, mimicked by a constant and homogeneous force density $f = p'$.
In the absence of particles, this would lead to a parabolic velocity profile with peak velocity $\hat{u}_0 = p' D^2 / (16 \eta_\text{ex})$, average velocity $\bar{u}_0 = \hat{u}_0 / 2$ and wall shear rate $\WSR = 4 \hat{u}_0 / D$.
RBCs in such a flow experience viscous shear stresses that can be characterised by the capillary number
\begin{equation}
 \label{eq:capillarynumber}
 \Canum = \frac{\eta_\text{ex} \WSR \rRBC}{\kappa_\text{S}} = \frac{p' D \rRBC}{4 \kappa_\text{S}}.
\end{equation}
Under physiological conditions ($\eta_\text{ex}$, $\kappa_\text{S}$ and $\rRBC$ as reported in \tabref{tab:symbols}), $\Canum = 1$ corresponds to $\WSR \approx 1300\,\text{s}^{-1}$, which leads to significant RBC deformations.
Note that the definition of $\Canum$ in Poiseuille flow is not unique and that there may be a constant conversion factor between the present and other authors' choice.

The purpose of this article is to investigate the effect of tube diameter $D$ and capillary number $\Canum$ on the platelet dynamics in tube flow at physiological blood parameters.
\tabref{tab:parameters} lists the investigated values for $D$ and $\Canum$.
The tube haematocrit is kept at $\Htt \approx 37\%$ in all cases.

In order to find a dimensionless time for the \emph{advection} of the suspension, one may define the advection time scale
\begin{equation}
 \label{eq:timescale}
 t_\text{ad} = \frac{2 \rRBC}{\bar{u}_0}.
\end{equation}
It is the average time an RBC requires to move its own diameter in the unperturbed flow.

\subsection{Simulation parameters}
\label{sec:simulationparameters}

\begin{table}[b]
 \caption{Overview of numerical simulation parameters. [s.u.] denotes ``simulation units''. Symbols are explained in \tabref{tab:symbols}. Other simulation parameters that are the same throughout are $\nu_\text{in} = 5/6$, $\nu_\text{ext} = 1/6$, $\Htt \approx 37\%$, $\kappa_\alpha = 0.5$, $\kappa_\text{A} = \kappa_\text{V} = 1$ and $N_t = 2 \cdot 10^5$.}
 \label{tab:parameters}
 \centering
 \begin{tabular}{llllll}
  \hline\noalign{\smallskip}
  $D$ & $\Canum$ & $N_\text{pl}$ & $p'$ & $\kappa_\text{S}$ & $\kappa_\text{B}$ \\
  $[\micron]$ & & & [s.u.] & [s.u.] & [s.u.]\\
  \noalign{\smallskip}\hline\noalign{\smallskip}
  \multirow{6}{*}{$10$} & $0.1$ & \multirow{6}{*}{$7$} & \multirow{6}{*}{$1.48 \cdot 10^{-4}$} & $0.133$ & $0.0453$ \\
  & $0.2$ & & & $0.0666$ & $0.0226$ \\
  & $0.3$ & & & $0.0444$ & $0.0151$ \\
  & $0.6$ & & & $0.0222$ & $0.00755$ \\
  & $1.0$ & & & $0.00453$ & $0.00453$ \\
  & $2.0$ & & & $0.00666$ & $0.00226$ \\ \hline
  \multirow{6}{*}{$15$} & $0.1$ & \multirow{6}{*}{$16$} & \multirow{6}{*}{$6.58 \cdot 10^{-5}$} & $0.0888$ & $0.0302$ \\
  & $0.2$ & & & $0.0444$ & $0.0151$ \\
  & $0.3$ & & & $0.0296$ & $0.0101$ \\
  & $0.6$ & & & $0.0148$ & $0.00503$ \\
  & $1.0$ & & & $0.00888$ & $0.00302$ \\
  & $2.0$ & & & $0.00444$ & $0.00151$ \\ \hline
  \multirow{6}{*}{$20$} & $0.1$ & \multirow{6}{*}{$28$} & \multirow{6}{*}{$3.70 \cdot 10^{-5}$} & $0.0666$ & $0.0226$ \\
  & $0.2$ & & & $0.0333$ & $0.0113$ \\
  & $0.3$ & & & $0.0222$ & $0.00755$ \\
  & $0.6$ & & & $0.0111$ & $0.00377$ \\
  & $1.0$ & & & $0.00666$ & $0.00226$ \\
  & $2.0$ & & & $0.00333$ & $0.00113$ \\ \hline
  \multirow{6}{*}{$30$} & $0.1$ & \multirow{6}{*}{$63$} & \multirow{6}{*}{$1.65 \cdot 10^{-5}$} & $0.0444$ & $0.0151$ \\
  & $0.2$ & & & $0.0222$ & $0.00755$ \\
  & $0.3$ & & & $0.0148$ & $0.00503$ \\
  & $0.6$ & & & $0.00741$ & $0.00252$ \\
  & $1.0$ & & & $0.00444$ & $0.00151$ \\
  & $2.0$ & & & $0.00222$ & $0.000755$ \\
  \noalign{\smallskip}\hline
  \noalign{\smallskip}\hline
 \end{tabular}
\end{table}

The flow is periodic along the tube axis ($x$-direction).
The circular tube cross-section is approximated by a staircase, and the half-way bounce-back boundary condition is used to enfore no-slip at the wall \citep{ladd_numerical_1994}.
Due to the large tube diameters and the choice of the external viscosity, artificial numerical slip is negligible.

The lattice resolution is $\Delta x = 0.33\, \micron$, \emph{i.e.}~the RBC diameter is $2 \rRBC = 8\, \micron = 24 \Delta x$.
While the tube diameter $D$ varies, the length of the simulated tube segment is $L = 12 \rRBC = 144 \Delta x$ in all simulations.

Obeying $\Lambda = 5$, the viscosities are $\nu_\text{ex} = 1/6$ and $\nu_\text{in} = 5/6$ in simulation units.
This leads to BGK relaxation parameters $\tau_\text{ex} = 1$ and $\tau_\text{in} = 3$, according to Eq.~\eqref{eq:lbm_viscosity}.

The flow is driven by a constant force density $f = p'$ along the tube axis.
Its value is chosen in such a way that the centre velocity $\hat{u}_0$ would be $0.05\, \Delta x / \Delta t$ in the absence of any particles.
This is to avoid compressibility effects and to keep the time step sufficiently small to achieve stable simulations.
Note that the actual centre velocity is smaller (typically by a factor of $\approx 2$) due to the presence of the cells.

\tabref{tab:parameters} lists all relevant parameter values for the simulations undertaken.

\subsection{Simulation initialisation}
\label{sec:initialization}

Due to the relatively small system size (between $N_\text{RBC} = 14$ and $126$), the number of platelets has been taken as $N_\text{pl} = N_\text{RBC} / 2$.
This is about $7$--$8$ times larger than observed under physiological conditions.
Since platelets are much smaller than RBCs, their volume fraction is still small compared to $\Htt$.

At the beginning of a simulation, platelets and RBCs are distributed throughout the tube with random positions and orientations.
Any overlap of particles with particles and particles with the tube wall is avoided.
The platelets are positioned first, which assures a more homogeneous distribution of them across the tube cross-section.
In order to facilitate this procedure, all particles are initially shrunk to half of their linear size and grown afterwards within 4000 time steps.
During growth, the particle volume is increased with a constant rate and overlap is avoided by a repulsion force.

As the growth process leads to an increase of kinetic energy in the system, a friction force is used to dissipate energy.
This force is switched off once the growth process is complete.
The exact form of the dissipation force is not significant for this purpose.

After all particles have reached their full size, the simulation starts with a constant fluid density $\rho$ (unity in simulation units) and zero velocity everywhere.
The force density $f = p'$ along the tube is switched on instantaneously to drive the flow.
Each simulation runs for $N_t = 2 \cdot 10^5$ time steps, i.e.~about $208\, \tad$.

\section{Simulation results and discussion}
\label{sec:results}

The results and discussion are presented in three parts.
In section~\ref{subsec:visual}, the final suspension state is visually inspected and characterised, and the cell-free layer thickness is investigated.
Section~\ref{subsec:margination} contains the margination observations and analysis.
Platelet dynamics near the tube wall is discussed in section~\ref{subsec:tumbling}.

\subsection{Visual results and cell-free layer thickness}
\label{subsec:visual}

\begin{figure*}[t]
 \centering
 \subfigure[\label{subfig:configuration_a} $D = 10\, \micron$, $\text{Ca} = 0.1$]{\includegraphics[width=0.475\linewidth]{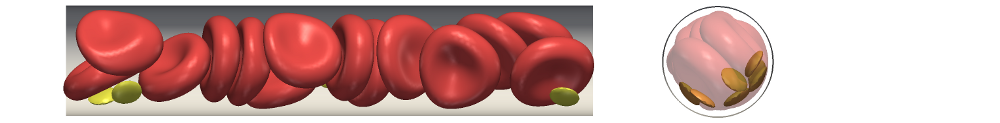}} \hfill
 \subfigure[\label{subfig:configuration_b} $D = 10\, \micron$, $\text{Ca} = 2.0$]{\includegraphics[width=0.475\linewidth]{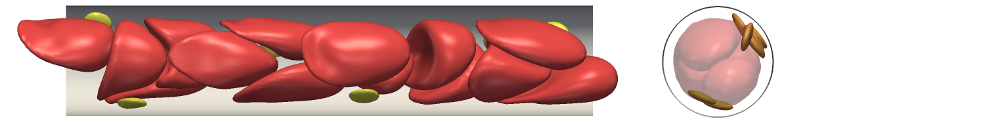}} \\
 \subfigure[\label{subfig:configuration_c} $D = 15\, \micron$, $\text{Ca} = 0.1$]{\includegraphics[width=0.475\linewidth]{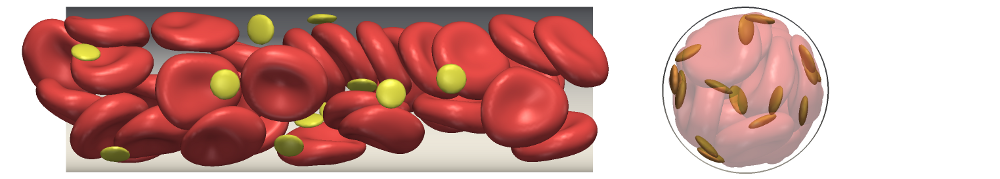}} \hfill
 \subfigure[\label{subfig:configuration_d} $D = 15\, \micron$, $\text{Ca} = 2.0$]{\includegraphics[width=0.475\linewidth]{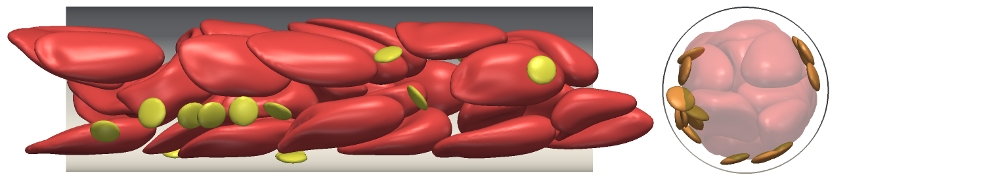}} \\
 \subfigure[\label{subfig:configuration_e} $D = 20\, \micron$, $\text{Ca} = 0.1$]{\includegraphics[width=0.475\linewidth]{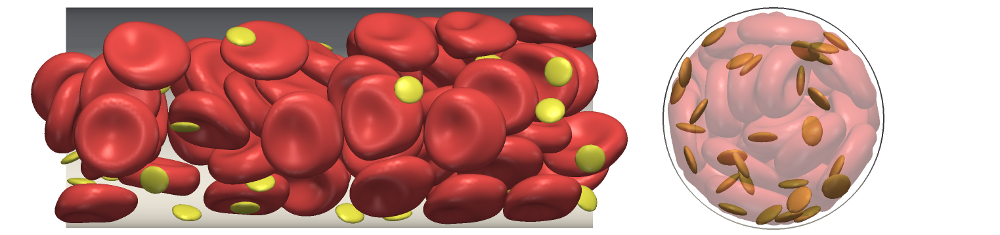}} \hfill
 \subfigure[\label{subfig:configuration_f} $D = 20\, \micron$, $\text{Ca} = 2.0$]{\includegraphics[width=0.475\linewidth]{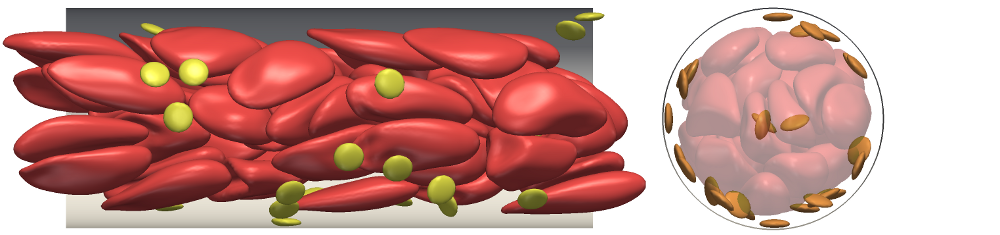}} \\
 \subfigure[\label{subfig:configuration_g} $D = 30\, \micron$, $\text{Ca} = 0.1$]{\includegraphics[width=0.475\linewidth]{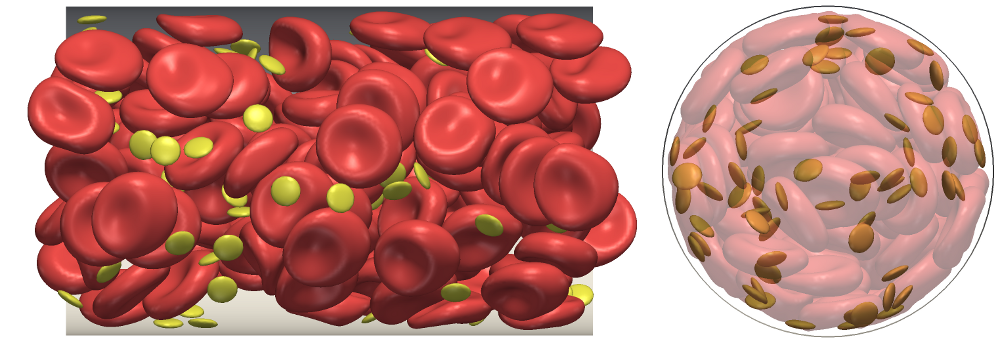}} \hfill
 \subfigure[\label{subfig:configuration_h} $D = 30\, \micron$, $\text{Ca} = 2.0$]{\includegraphics[width=0.475\linewidth]{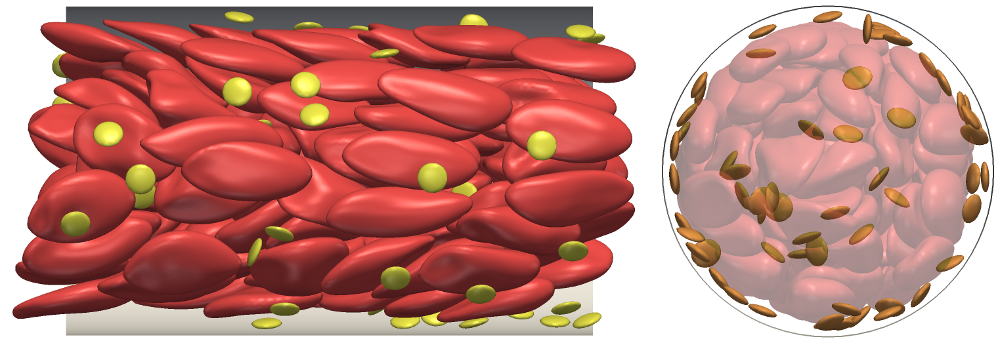}} \\
 \caption{Configuration of the red blood cells (red) and platelets (yellow) for all investigated tube diameters ($D = 10$, $15$, $20$, $30\, \micron$) and the smallest and largest studied capillary numbers ($\Canum = 0.1$, $2.0$) at $\Htt \approx 37\%$ after $t = 208\, t_\text{ad}$. The left columns show sections along the tube, the right columns tube cross-sections with reduced RBC opacity to reveal all platelets. Flow is from left to right (left columns) and out of the image plane (right columns).}
 \label{fig:configuration}
\end{figure*}

\begin{figure*}[t]
 \subfigure[\label{subfig:CFL1} Radial haematocrit distribution ($D = 30\, \micron$, $\Canum = 2.0$)]{\includegraphics[width=0.475\linewidth]{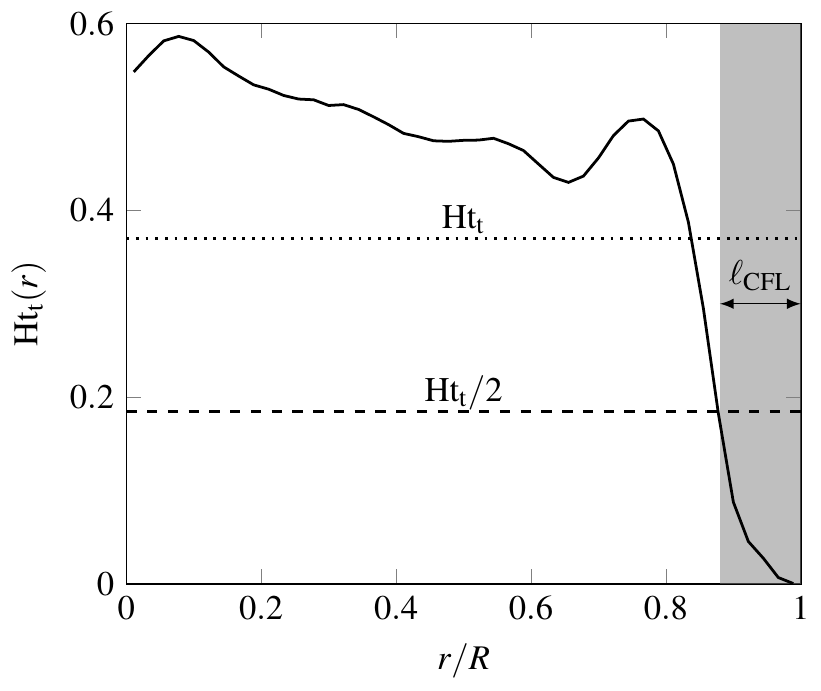}} \hfill
 \subfigure[\label{subfig:CFL2} CFL thickness]{\includegraphics[width=0.475\linewidth]{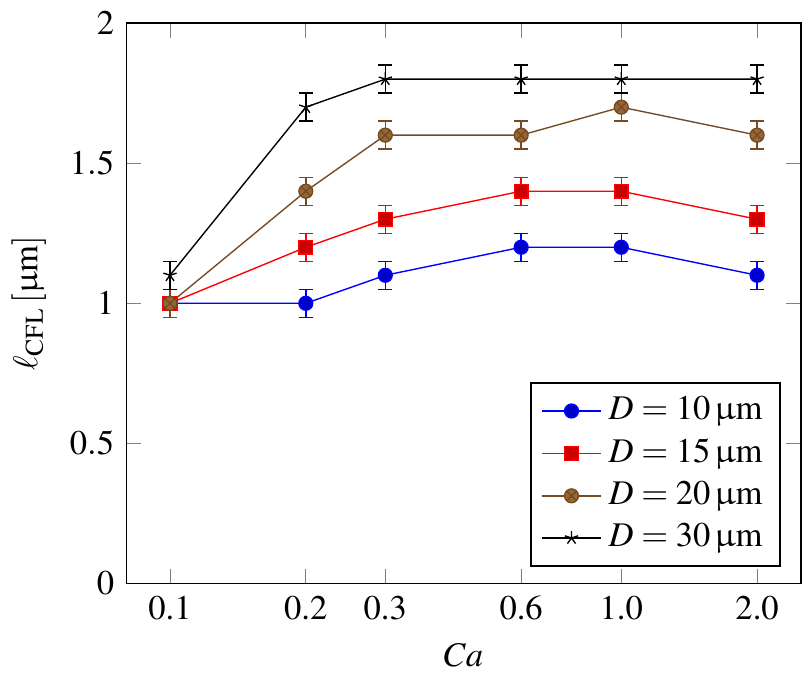}} \\
 \caption{Cell-free layer thickness $\CFL$. (a) Exemplary radial tube haematocrit profile $\Htt(r)$ averaged over time. The CFL thickness $\CFL$ (denoted by grey area) is defined as the distance from the wall ($r = R$) to the point where $\Htt(r)$ reaches half of the average tube haematocrit. (b) $\CFL$ as function of capillary number $\Canum$ for all investigated tube diameters $D$ at $\Htt \approx 37\%$. Error bars are estimated. Lines are guides for the eye. Note that the $\Canum$-axis is logarithmic.}
 \label{fig:CFL}
\end{figure*}

Fig.~\ref{fig:configuration} shows the final state (after 208 advection times) for the smallest and largest simulated capillary number $\Canum$ and each tube diameter $D$.
Both the RBC and platelet configurations depend on the geometry and flow parameters.

At $\Canum = 0.1$, RBCs are only weakly deformed, and their dimples are easy to spot.
Contrarily, at $\Canum = 2.0$, RBCs show a strong and characteristic elongation, and the dimples have disappeared.
The CFL is particulary pronounced at higher $\Canum$, and RBCs close to the tube wall tend to align with the wall and form a ring.
This leads to a slight increase in the local RBC concentration (Fig.~\ref{subfig:CFL1}) as observed previously \citep{lei_blood_2013}.
Furthermore, RBCs in Fig.~\ref{subfig:configuration_b} show a typical zigzag configuration that can be found in small blood vessels \citep{mcwhirter_flow-induced_2009}.

The platelet distribution changes strongly with $D$ and $\Canum$, as discussed in more detail in section \ref{subsec:margination}.
In summary, platelets prefer to marginate towards the tube wall with increasing $\Canum$.
Also, for smaller $D$, a larger fraction of platelets resides near the tube wall at the end of the simulations.

There is no unique way to define the CFL thickness $\CFL$ since the edge between the outermost RBCs and the plasma layer is diffuse \citep{lei_blood_2013}.
The blurring of the CFL is clearly visible in Fig.~\ref{fig:configuration}.
In the following, the definition of $\CFL$ is based on the radial tube haematocrit profile $\Htt(r)$ averaged over time.
As shown in Fig.~\ref{subfig:CFL1}, $\Htt(r)$ decreases strongly near the wall.
The CFL thickness is defined by the distance from the tube wall where $\Htt(r)$ reaches $\Htt/2$.

Fig.~\ref{subfig:CFL2} shows the resulting CFL thickness as function of $\Canum$ for different tube diameters $D$.
$\CFL$ is generally increasing with $\Canum$ until it saturates above $\Canum \approx 0.3$--$0.6$.
This is close to the point where RBC tumbling is replaced by tank-treading \citep{kruger_crossover_2013}.
The slight decrease of $\CFL$ at $\Canum = 2$ is caused by a subtle geometrical rearrangement of the RBCs due to their large deformations.
Evidently, the CFL thickness is larger for wider tubes, but the relative thickness $\CFL / D$ is decreasing.
In the macroscopic limit, for $D > 100\,\micron$, the CFL is less important for the effective blood rheology \citep{lei_blood_2013}.

In the following, $\CFL$ is used to identify those platelets that are located between the RBC-rich region and the tube wall.

\subsection{Platelet margination}
\label{subsec:margination}

\begin{figure*}
 \subfigure[$D = 10\, \micron$, $\Canum = 0.1$]{\includegraphics[width=0.475\linewidth]{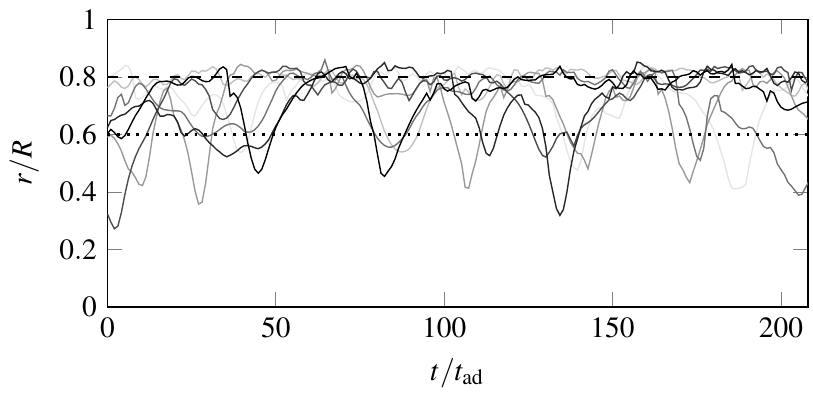}} \hfill
 \subfigure[$D = 10\, \micron$, $\Canum = 2.0$]{\includegraphics[width=0.475\linewidth]{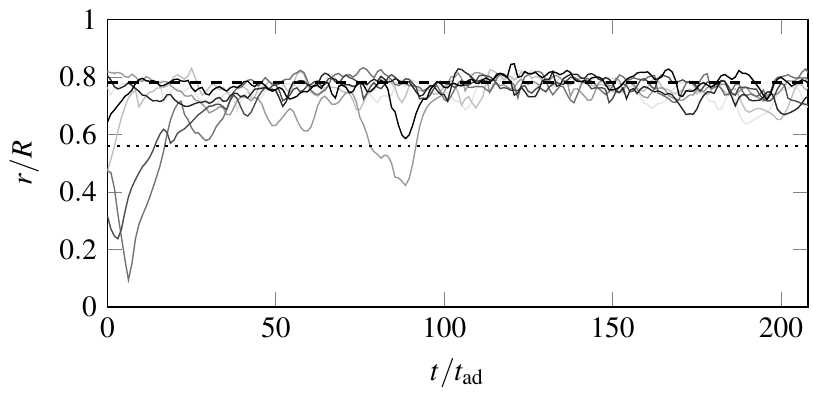}} \\
 \subfigure[$D = 15\, \micron$, $\Canum = 0.1$]{\includegraphics[width=0.475\linewidth]{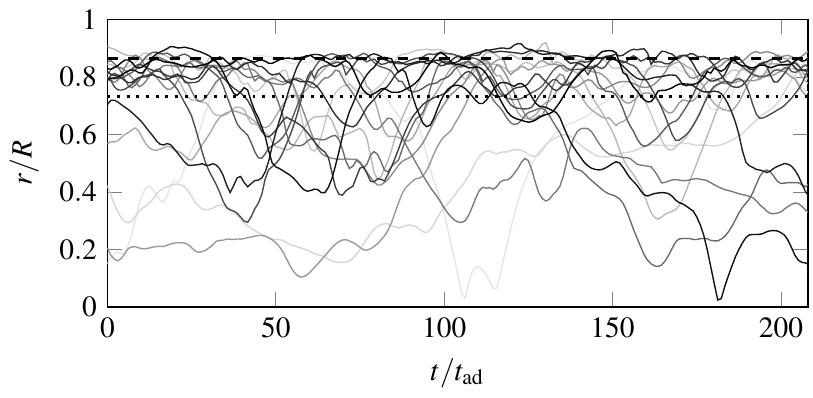}} \hfill
 \subfigure[$D = 15\, \micron$, $\Canum = 2.0$]{\includegraphics[width=0.475\linewidth]{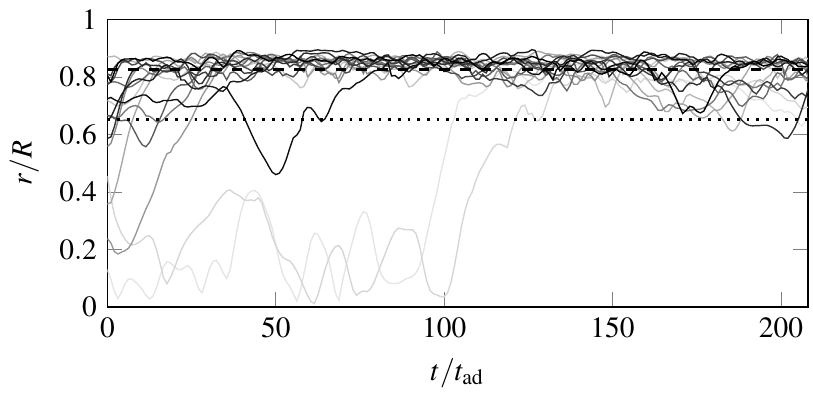}} \\
 \subfigure[$D = 20\, \micron$, $\Canum = 0.1$]{\includegraphics[width=0.475\linewidth]{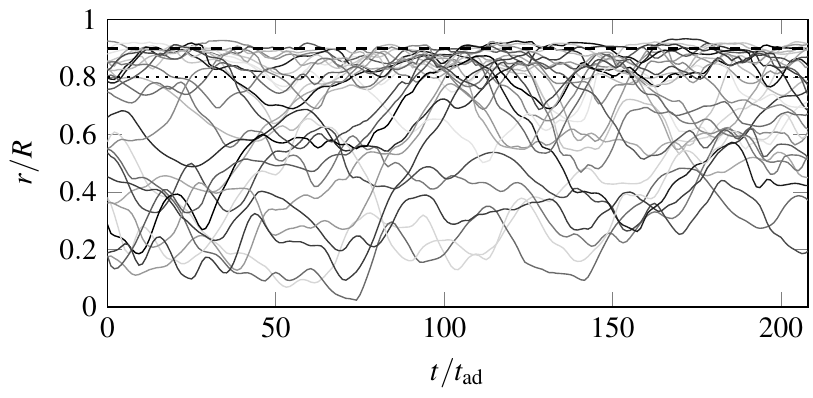}} \hfill
 \subfigure[$D = 20\, \micron$, $\Canum = 2.0$]{\includegraphics[width=0.475\linewidth]{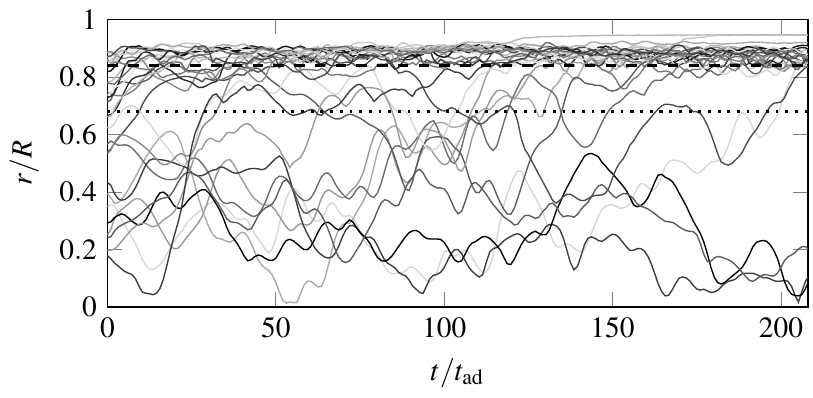}} \\
 \subfigure[$D = 30\, \micron$, $\Canum = 0.1$]{\includegraphics[width=0.475\linewidth]{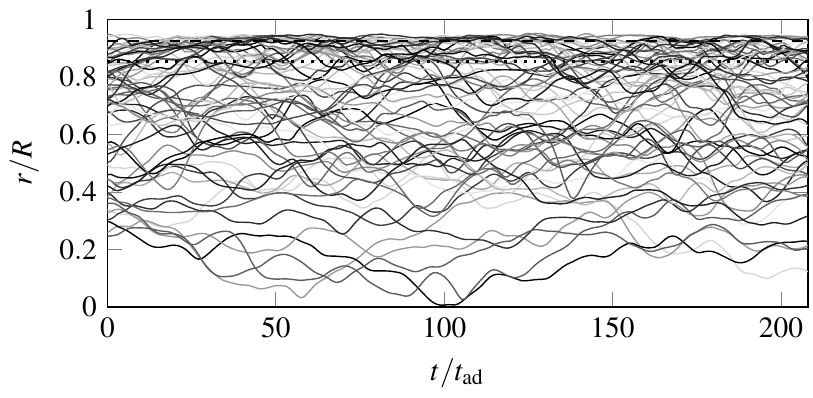}} \hfill
 \subfigure[$D = 30\, \micron$, $\Canum = 2.0$]{\includegraphics[width=0.475\linewidth]{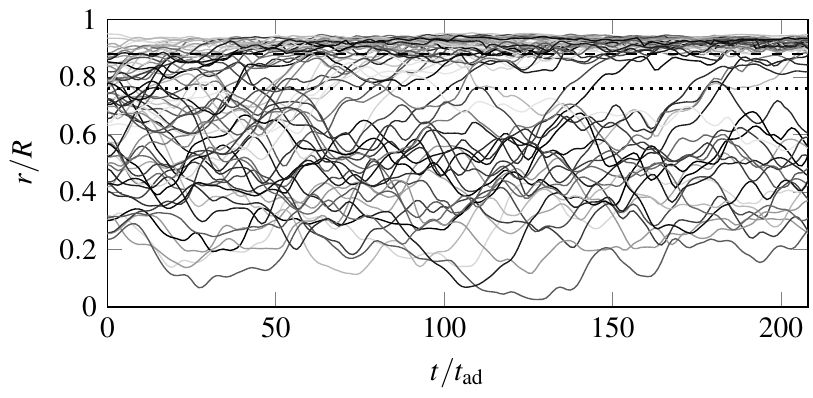}} \\
 \caption{Platelet margination for different tube diameters ($D = 10\, \micron$ with 7 platelets, $D = 15\, \micron$ with 16 platelets, $D = 20\, \micron$ with 28 platelets and $D = 30\, \micron$ with 63 platelets). The radial position $r$ (normalised by tube radius $R$) for each platelet is shown for two different capillary numbers ($\Canum = 0.1$ and $2.0$) as function of dimensionless time. The dashed line denotes the cell-free layer thickness with distance $\CFL$ from the wall. The dotted line denotes twice the CFL thickness, $2 \CFL$. Note that platelets are randomly distributed over the pipe cross-section at the start of the simulation. This means that more platelets are initially located at larger radii $r$ since the cross-section area element obeys $\text{d}A = 2 \pi r\, \text{d}r$.}
 \label{fig:margination}
\end{figure*}

\begin{figure*}
 \subfigure[$D = 10\, \micron$, $\Canum = 0.1$]{\includegraphics[width=0.3\linewidth]{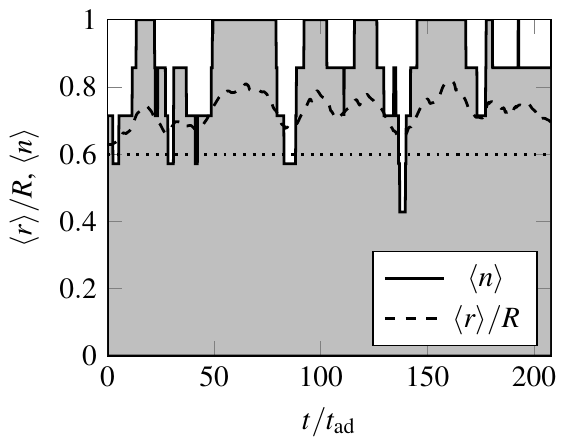}} \hfill
 \subfigure[$D = 10\, \micron$, $\Canum = 0.3$]{\includegraphics[width=0.3\linewidth]{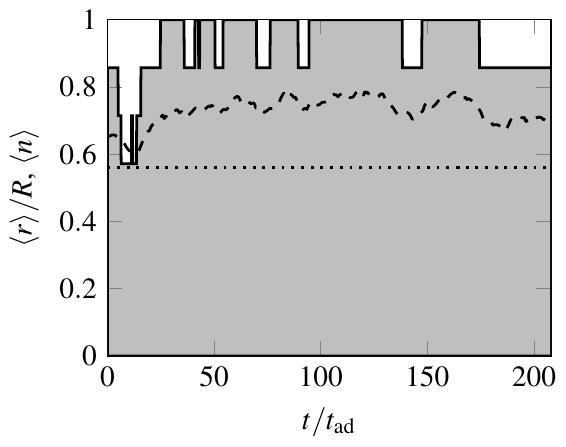}} \hfill
 \subfigure[$D = 10\, \micron$, $\Canum = 2.0$]{\includegraphics[width=0.3\linewidth]{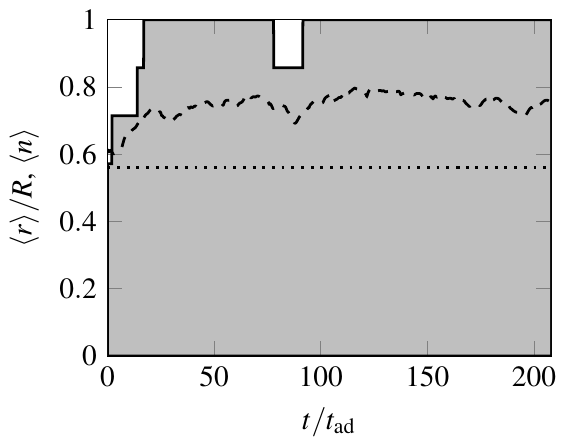}} \\
 \subfigure[$D = 15\, \micron$, $\Canum = 0.1$]{\includegraphics[width=0.3\linewidth]{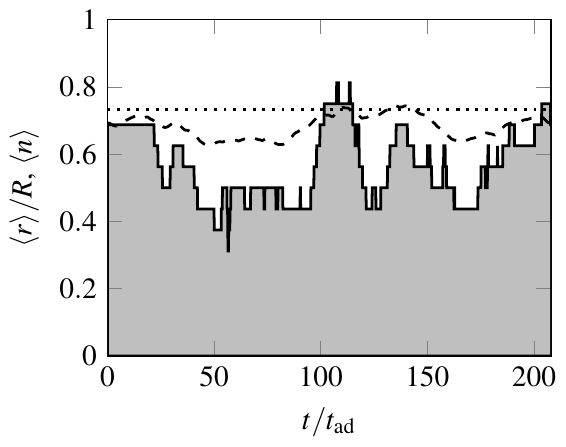}} \hfill
 \subfigure[$D = 15\, \micron$, $\Canum = 0.3$]{\includegraphics[width=0.3\linewidth]{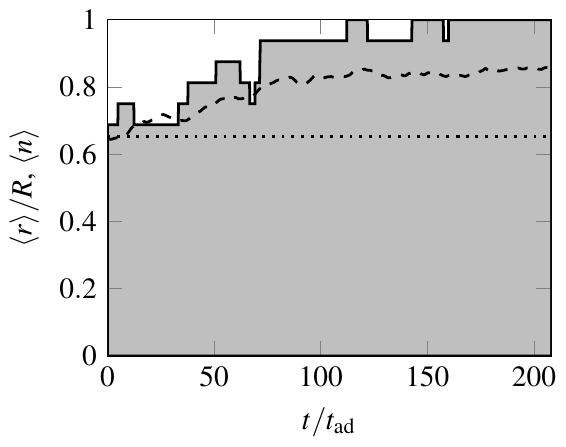}} \hfill
 \subfigure[$D = 15\, \micron$, $\Canum = 2.0$]{\includegraphics[width=0.3\linewidth]{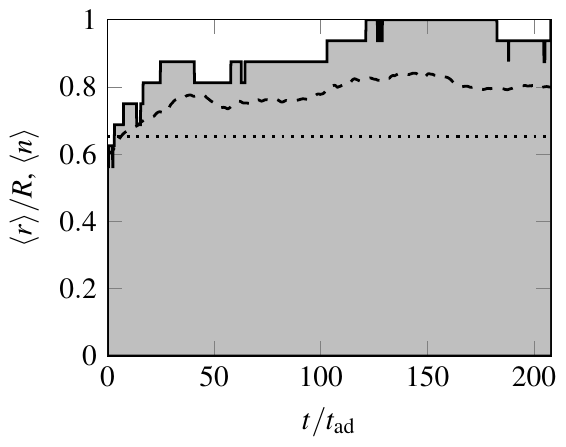}} \\
 \subfigure[$D = 20\, \micron$, $\Canum = 0.1$]{\includegraphics[width=0.3\linewidth]{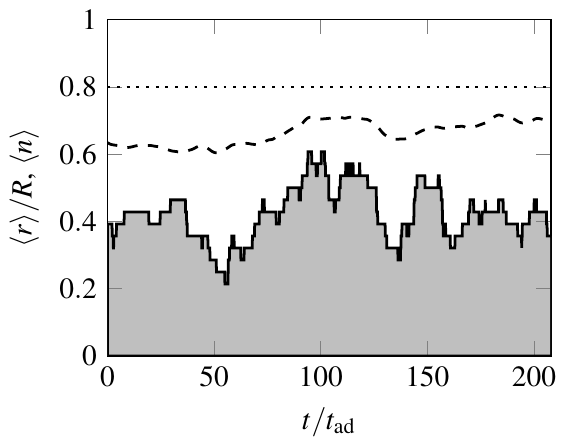}} \hfill
 \subfigure[$D = 20\, \micron$, $\Canum = 0.3$]{\includegraphics[width=0.3\linewidth]{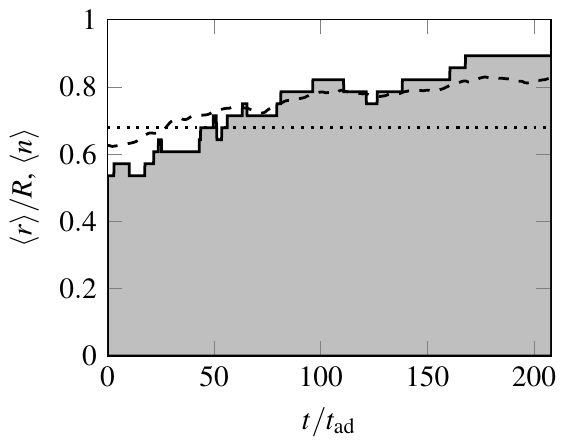}} \hfill
 \subfigure[$D = 20\, \micron$, $\Canum = 2.0$]{\includegraphics[width=0.3\linewidth]{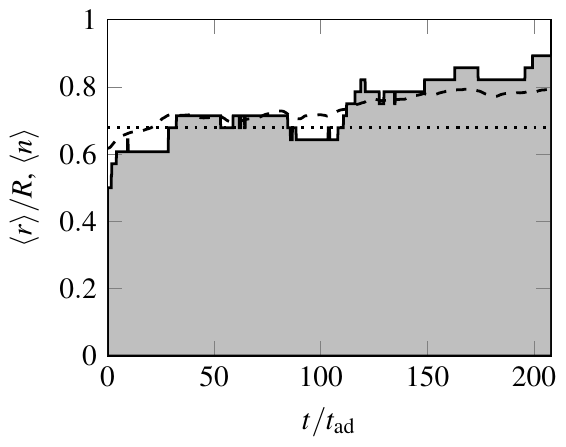}} \\
 \subfigure[$D = 30\, \micron$, $\Canum = 0.1$]{\includegraphics[width=0.3\linewidth]{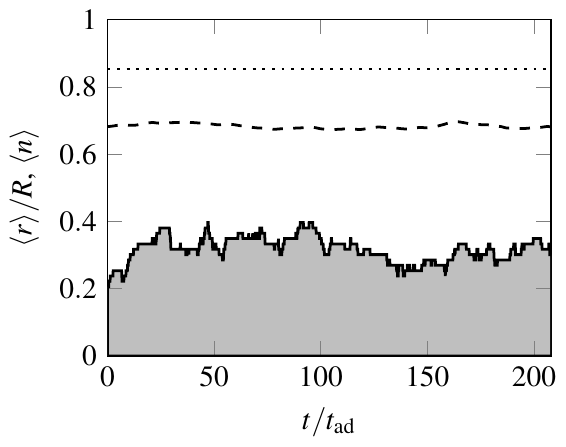}} \hfill
 \subfigure[$D = 30\, \micron$, $\Canum = 0.3$]{\includegraphics[width=0.3\linewidth]{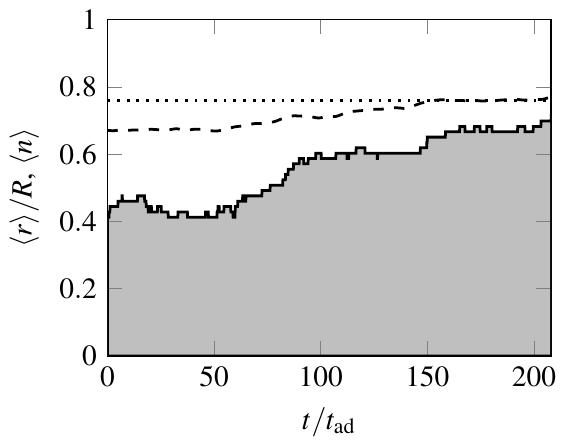}} \hfill
 \subfigure[$D = 30\, \micron$, $\Canum = 2.0$]{\includegraphics[width=0.3\linewidth]{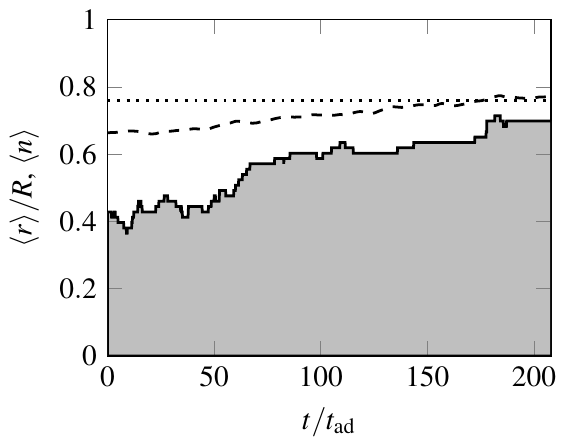}} \\
 \caption{Average platelet margination for all studied tube diameters $D$ (varied row-wise) and a few capillary numbers ($\Canum = 0.1$, $0.3$, $2.0$, varied column-wise). Dashed lines show the radial position $\langle r \rangle$ averaged over all platelets as function of time. The grey regions bounded by solid lines indicate the number fraction $\langle n \rangle$ of all platelets that are located within a region $2 \CFL$ from the tube wall (indicated by horizontal dotted lines).}
 \label{fig:margination_radius}
\end{figure*}

\begin{figure*}
 \subfigure[$D = 10\, \micron$, $\Canum = 0.1$]{\includegraphics[width=0.3\linewidth]{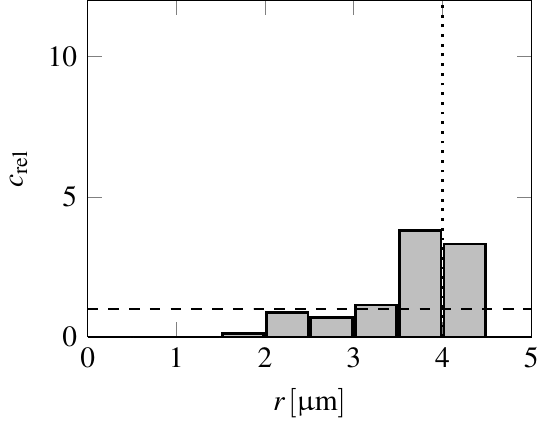}} \hfill
 \subfigure[$D = 10\, \micron$, $\Canum = 0.3$]{\includegraphics[width=0.3\linewidth]{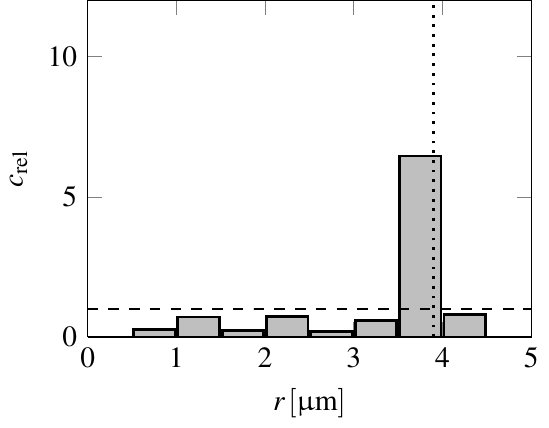}} \hfill
 \subfigure[$D = 10\, \micron$, $\Canum = 2.0$]{\includegraphics[width=0.3\linewidth]{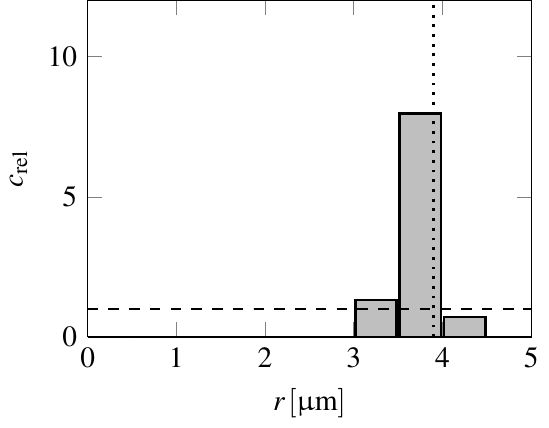}} \\
 \subfigure[$D = 15\, \micron$, $\Canum = 0.1$]{\includegraphics[width=0.3\linewidth]{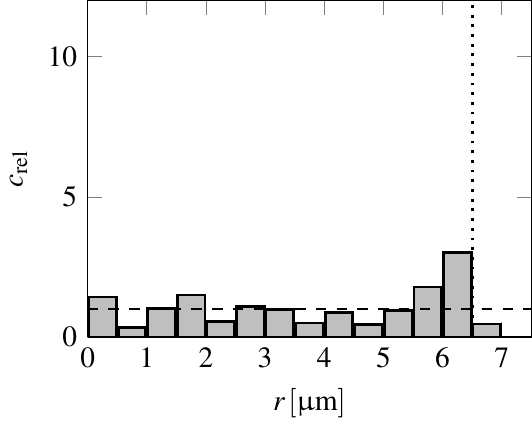}} \hfill
 \subfigure[$D = 15\, \micron$, $\Canum = 0.3$]{\includegraphics[width=0.3\linewidth]{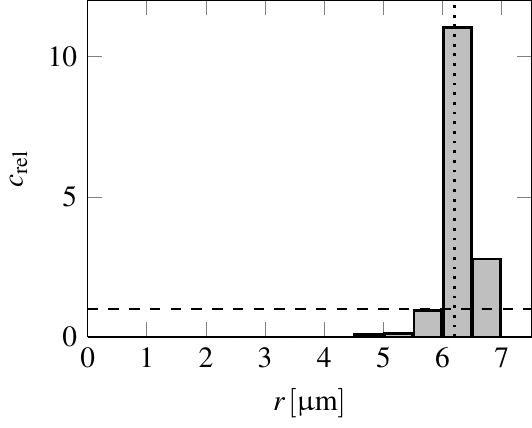}} \hfill
 \subfigure[$D = 15\, \micron$, $\Canum = 2.0$]{\includegraphics[width=0.3\linewidth]{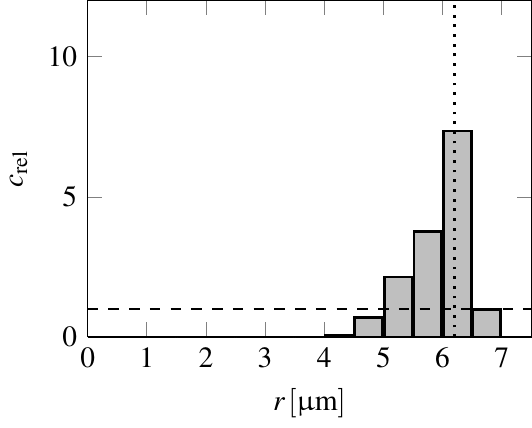}} \\
 \subfigure[$D = 20\, \micron$, $\Canum = 0.1$]{\includegraphics[width=0.3\linewidth]{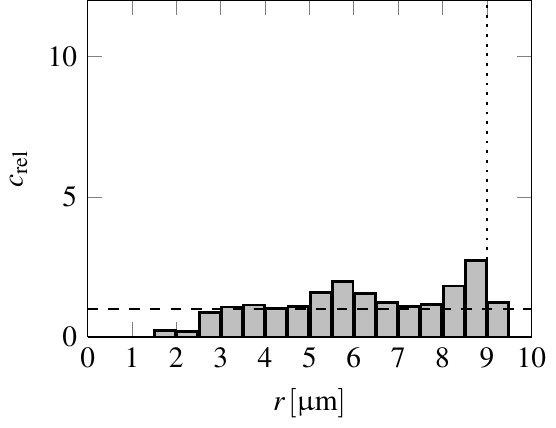}} \hfill
 \subfigure[$D = 20\, \micron$, $\Canum = 0.3$]{\includegraphics[width=0.3\linewidth]{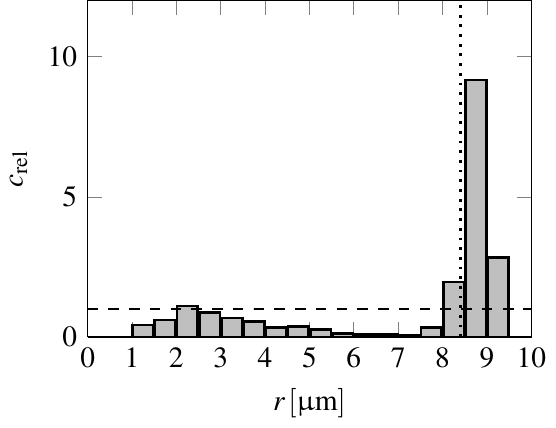}} \hfill
 \subfigure[$D = 20\, \micron$, $\Canum = 2.0$]{\includegraphics[width=0.3\linewidth]{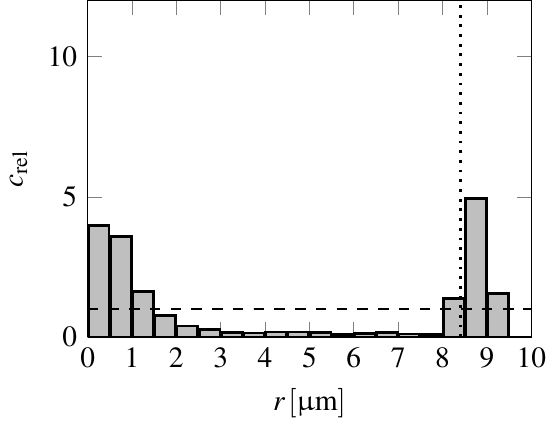}} \\
 \subfigure[$D = 30\, \micron$, $\Canum = 0.1$]{\includegraphics[width=0.3\linewidth]{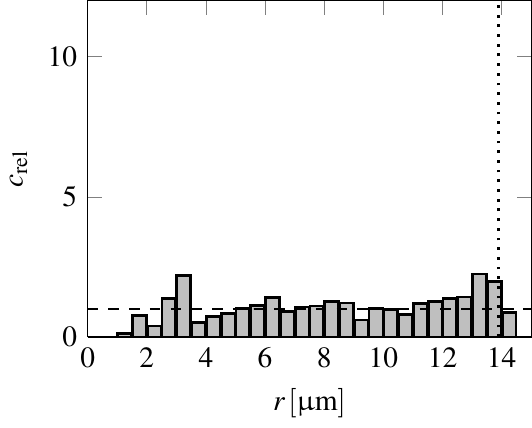}} \hfill
 \subfigure[$D = 30\, \micron$, $\Canum = 0.3$]{\includegraphics[width=0.3\linewidth]{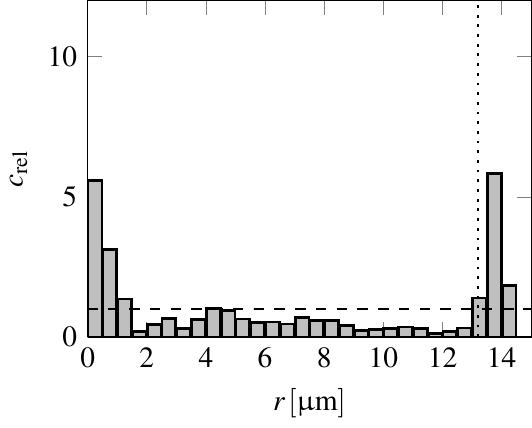}} \hfill
 \subfigure[$D = 30\, \micron$, $\Canum = 2.0$]{\includegraphics[width=0.3\linewidth]{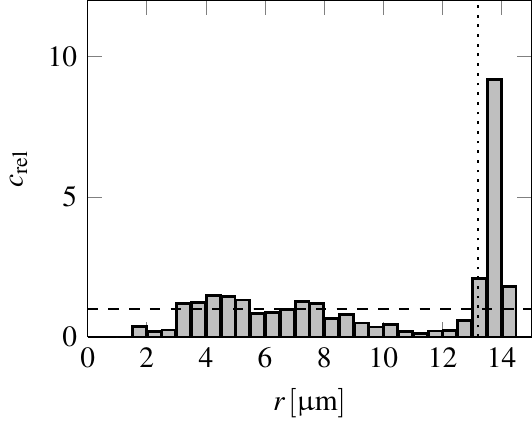}} \\
 \caption{Radial distribution of relative platelet concentration $c_\text{ref}$ for different tube diameters $D$ (varying row-wise) and capillary numbers $\Canum$ (varying column-wise). The platelet distributions are normalised by the cross-sectional area of each bin such that a homogeneous distribution would lead to a constant line at $c_\text{rel} = 1$ (denoted by horizontal dashed line). The bin size is $0.5\,\micron$. Data is obtained from platelet centres and averaged between $3 N_t / 4$ and $N_t$. The cell-free layer thickness is shown as vertical dotted line.}
 \label{fig:margination_bins}
\end{figure*}

\begin{figure*}
 \subfigure[radial average velocity ($D = 20\, \micron$)]{\includegraphics[width=0.475\linewidth]{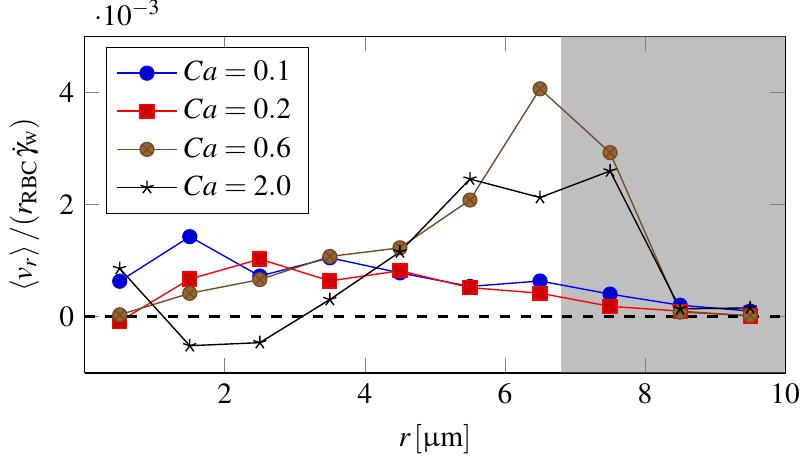}} \hfill
 \subfigure[radial velocity fluctuation ($D = 20\, \micron$)]{\includegraphics[width=0.475\linewidth]{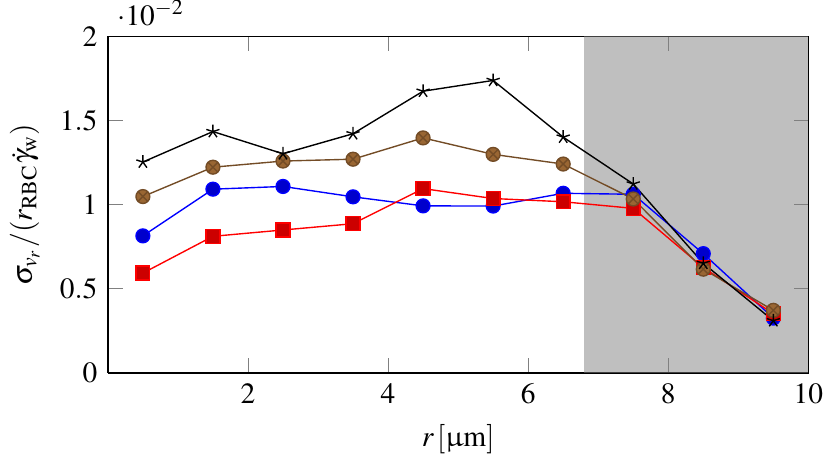}} \\
 \subfigure[radial average velocity ($D = 30\, \micron$)]{\includegraphics[width=0.475\linewidth]{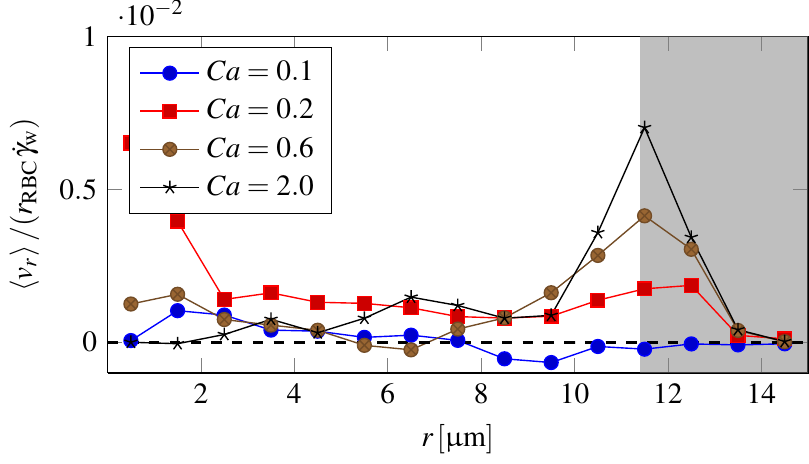}} \hfill
 \subfigure[radial velocity fluctuation ($D = 30\, \micron$)]{\includegraphics[width=0.475\linewidth]{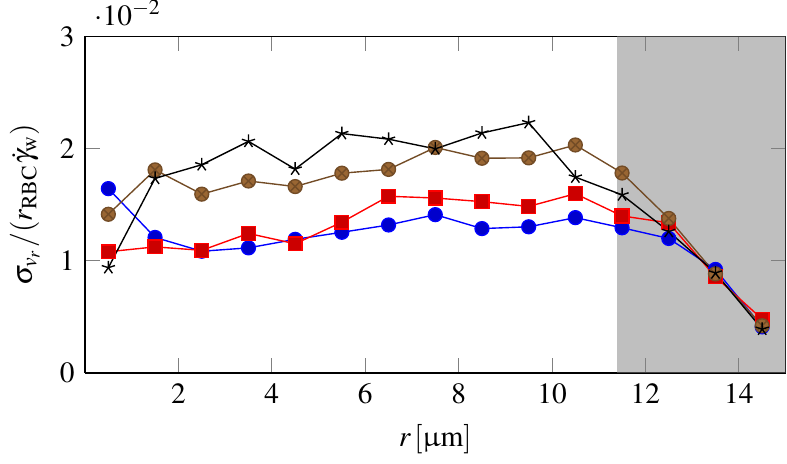}} \\
 \caption{(a,c) Average radial platelet velocity $\left \langle v_r \right \rangle$ and (b,d) its standard deviation $\sigma_{v_r}$ as function of radial position $r$. Data is shown for (a,b) $D = 20\,\micron$ and (c,d) $D = 30\,\micron$. Data is time-averaged after the first quarter of the simulation (i.e.~between $\approx 50$ and $200$ advection times), and velocities are normalised by $\rRBC \WSR$. The grey-shaded area denotes the CFL vicinity ($2 \CFL$). Lines are guides for the eyes.}
 \label{fig:radialvelocity}
\end{figure*}

Some exemplary radial platelet positions as function of time are shown in Fig.~\ref{fig:margination}.
In the following, a platelet is considered being fully marginated when it is located in the CFL vicinity with its radial centre position between $R - 2 \CFL$ and $R$ (indicated by dotted lines in Fig.~\ref{fig:margination}).
As seen in Fig.~\ref{subfig:CFL1}, this corresponds to the region between the tube wall and the haematocrit peak at the CFL edge.
In this region, $\Htt(r)$ is a strongly decreasing function.

The character of the platelet motion depends on the tube diameter $D$ and the capillary number $\Canum$.
\cite{zhao_shear-induced_2011} demonstrated previously that platelets show diffusive behaviour in the RBC-rich region until they reach the CFL where they get trapped irreversibly.
However, the present data shows that this is only the case for larger capillary numbers ($\Canum > 0.2$).
Fig.~\ref{fig:margination} reveals that, for $\Canum = 0.1$, several platelets leave the CFL vicinity again.
Those platelets do not move far away from the CFL, though; instead they tend to become caught again shortly after.
For larger $\Canum$, platelets are trapped in the CFL once they visit it for the first time.

Since it is difficult to extract further information directly from the trajectories in Fig.~\ref{fig:margination}, Fig.~\ref{fig:margination_radius} shows the time evolution of the average radial platelet position and the fraction of platelets that are located within $2 \CFL$ from the tube wall.
The first observation is that the time for platelets to become trapped increases with tube diameter.
In larger tubes, platelets have to move for a longer distance to reach the CFL; this takes more time.
From the third and fourth rows in Fig.~\ref{fig:margination_radius} one can conclude that margination is still ongoing for $D \geq 20\,\micron$ at the end of the simulation (after 208 advection times).
On the contrary, the platelet distributions in the smaller tubes with $D \leq 15\,\micron$ seem to have reached a quasi-equilibrium.

Secondly, margination and trapping are faster for higher capillary number.
However, there is no strong difference between $\Canum = 0.3$ and $\Canum = 2.0$ as the second and third columns in Fig.~\ref{fig:margination_radius} show.
\cite{zhao_shear-induced_2011} observed that the platelet diffusion in the RBC-rich region does not change significantly between $\Canum = 1.0$ and $\Canum = 2.0$, a fact that is attributed to the saturation of RBC deformation due to the conservation of the RBC surface area.
The strong increase of the margination efficiency between $\Canum = 0.1$ and $\Canum = 0.3$ in Fig.~\ref{fig:margination_radius} could be related to the dynamical state of the RBCs.
RBCs tend to tumble below and tank-tread above $\Canum \approx 0.2$ \citep{kruger_crossover_2013}.
\cite{reasor_determination_2013} hypothesised that tank-treading RBCs may increase the margination efficiency.
In fact, for $\Canum = 0.1$ (first column in Fig.~\ref{fig:margination_radius}), margination within the first 208 advection times is nearly absent, except for the smallest tube with $D = 10\,\micron$.

The smallest tube ($D = 10\, \micron$) plays a special role.
Platelets tend to reach the CFL nearly instantaneously, even at $\Canum = 0.1$.
Therefore, margination in small vessels with $D \leq 10\, \micron$ seems to be very effective for all capillary numbers.
This is in line with the observations by \cite{reasor_determination_2013} who reported that, for rigid RBCs, margination is only observed for platelets that are not farther away from the wall than about $5\, \micron$.
Therefore, all platelets in a $10\,\micron$-tube are always in wall vicinity.
This will be further discussed below.
Already for a slightly larger tube ($D = 15\, \micron$), several platelets remain in the RBC-rich region for an extended period of time if the RBCs are sufficiently rigid ($\Canum = 0.1$).
For tubes with $D = 20\, \micron$ and $30\, \micron$, only a fraction of the platelets initially located in the RBC-rich phase diffuse into the CFL during the simulation time.

As a final remark regarding Fig.~\ref{fig:margination_radius}, the number ratio of platelets near the tube wall fluctuates more for smaller $\Canum$.
This supports the observation that platelets are not irreversibly trapped in the CFL for $\Canum = 0.1$.
Further data substantiating this claim will be discussed shortly.

Fig.~\ref{fig:margination_bins} shows some radial platelet distributions averaged over the final $25\%$ of the simulation time (between $\approx 150$ and $200$ advection times).
These data support the previous interpretations.
For $D = 10\,\micron$ (first row), margination is very effective, and nearly all platelets are located close to the CFL (vertical dotted line).
The situation is different for larger tube diameters with $\Canum = 0.1$ (left column).
In those cases, there is nearly no margination, and the platelet distribution is close to being homogeneous (horizontal dashed line).
Much longer simulations are necessary to decide whether margination for small $\Canum$ is just slow or whether this is already the equilibrium state.
For $\Canum = 0.3$ and $\Canum = 2.0$ (second and third columns in Fig.~\ref{fig:margination_bins}), platelets are effectively transported towards the tube wall.

In Fig.~\ref{fig:margination_bins}(h,i,k,l), a dip in the platelet concentration close to the CFL edge is visible.
This suggests that platelets are effectively transported in outward direction when they are close to the CFL edge, therefore decreasing the concentration at $r \approx R - 2 \CFL$ and increasing it within the CFL.
Platelet diffusion leads to a fresh supply of platelets moving from the RBC-rich region towards the wall until there are no platelets left in the tube interior.
Further evidence supporting this interpretation will be discussed below.
As mentioned before, this process seems to have completed in Fig.~\ref{fig:margination_bins}(e,f), where the tube radius is relatively small ($D = 15\,\micron$).
Yet, the process appears to be still ongoing in Fig.~\ref{fig:margination_bins}(h,i,k,l), where the tube is larger.
In the latter cases, it is expected that all platelets will eventually reach the CFL after a sufficiently long time.

As a remark, there is also a platelet concentration peak near the centreline in Fig.~\ref{fig:margination_bins}(i,k).
Note that the absolute number of platelets in that region is small since the bin area decreases linearly with $r$ for $r \to 0$.
Therefore, the platelet concentration peak near the CFL edge is caused by many more platelets and is therefore more significant than the peak near $r = 0$.
This is nicely borne out in Fig.~\ref{subfig:configuration_f} where only three platelets are located near the centreline, but the remaining 25 are close to the CFL.
Therefore, the central peak is probably a statistical fluctuation due to the relatively short averaging period and limited platelet number.

To investigate the radial platelet transport in more detail, Fig.~\ref{fig:radialvelocity} shows the average radial platelet velocity $\left \langle v_r \right \rangle$ and its standard deviation $\sigma_{v_r}$ as function of radial position $r$ for $D = 20\,\micron$ and $D = 30\,\micron$.
Due to the small number of platelets and the fast margination, the radial velocity data for $D = 10\,\micron$ and $D = 15\,\micron$ is too noisy to provide useful information.

The first striking observation from Fig.~\ref{fig:radialvelocity} is that the radial platelet velocity is close to zero everywhere, except near the edge of the CFL (grey area) for $\Canum > 0.2$.
This shows that platelets are pushed towards the wall if they have reached the edge of the CFL.
This effect, however, only exists when RBCs are sufficiently strongly deformed and in the tank-treading state.
For the presented tube diameters, collisions between platelets and nearly rigid RBCs ($\Canum \leq 0.2$) do not lead to a significant lateral platelet transport near the CFL.
Note that the lateral platelet velocity is closely related to the drift term in effective platelet transport models \citep{eckstein_model_1991, yeh_transient_1994}.

The peak of $\left \langle v_r \right \rangle(r)$ coincides with the peak of $\Htt(r)$ in Fig.~\ref{subfig:CFL1}.
Therefore, platelets are already pushed outwards before they have passed the CFL edge.
For both tube diameters shown in Fig.~\ref{fig:radialvelocity}, $\left \langle v_r \right \rangle(r)$ becomes significantly larger than zero when the distance to the wall is about $5\,\micron$;
this is in good agreement with the results report by \cite{reasor_determination_2013}.

Fig.~\ref{fig:radialvelocity}(b,d) reveals that the radial platelet velocity fluctuation strongly decreases within the CFL.
This is an indicator for reduced platelet diffusion in that region.
One can see that the gradient of $\sigma_{v_r}$ at the edge of the CFL is larger for higher $\Canum$.
This suggests that escaping the CFL becomes increasingly difficult with growing $\Canum$ because it is less likely that a platelet is moving fast enough towards the tube axis before it is pushed back.
Fig.~\ref{fig:radialvelocity} can therefore explain why more platelets  are seen leaving the CFL again in Fig.~\ref{fig:margination} when $\Canum$ is small.

Furthermore, the radial velocity fluctuations in Fig.~\ref{fig:radialvelocity} in the RBC-rich region increase with $\Canum$.
This may result in a larger platelet diffusivity that further increases the lateral mobility of platelets when $\Canum$ is large.

Concluding this section, one can say that non-diffusive platelet margination in tubes with $D \geq 15\,\micron$ is only observed when $\Canum > 0.2$.
This is in good agreement with previous experimental results reporting that margination is only relevant when the wall shear rate is larger than $200\,\text{s}^{-1}$ \citep{tilles_near-wall_1987, eckstein_conditions_1988, bilsker_freeze-capture_1989}.
According to the definition of $\Canum$ in Eq.~\eqref{eq:capillarynumber}, $\Canum = 0.2$ corresponds to $\approx 250\,\text{s}^{-1}$.
This threshold value is close to the point where RBCs near the CFL edge are starting to tank-tread.
The onset of tank-treading therefore seems to be necessary for margination.
The situation is different for $D = 10\,\micron$ where platelets move to the tube wall very fast, even at $\Canum = 0.1$.
A possible explanation is that many RBCs in Fig.~\ref{subfig:configuration_a} fill the entire cross-section of the tube so that platelets can easily slide between them towards the CFL.

\subsection{Tumbling and sliding in the cell-free layer}
\label{subsec:tumbling}

\begin{figure}
 \includegraphics[width=\linewidth]{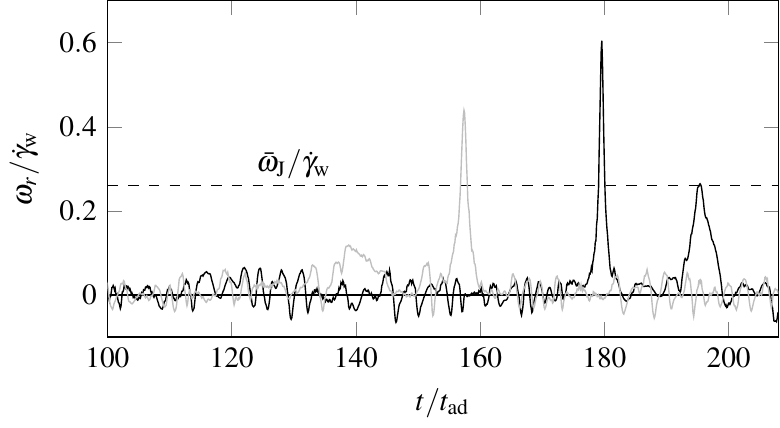}
 \caption{Angular velocity $\omega_r$ normalised by wall shear rate $\WSR$ of two platelets (grey and black) near the wall for $D = 20\,\micron$ and $\Canum = 0.6$. The dashed line indicates the expected average Jeffery frequency $\bar{\omega}_\text{J}$ for an unconfined platelet with aspect ratio $3.6$.}
 \label{fig:slidingtumbling}
\end{figure}

\begin{figure}
 \includegraphics[width=\linewidth]{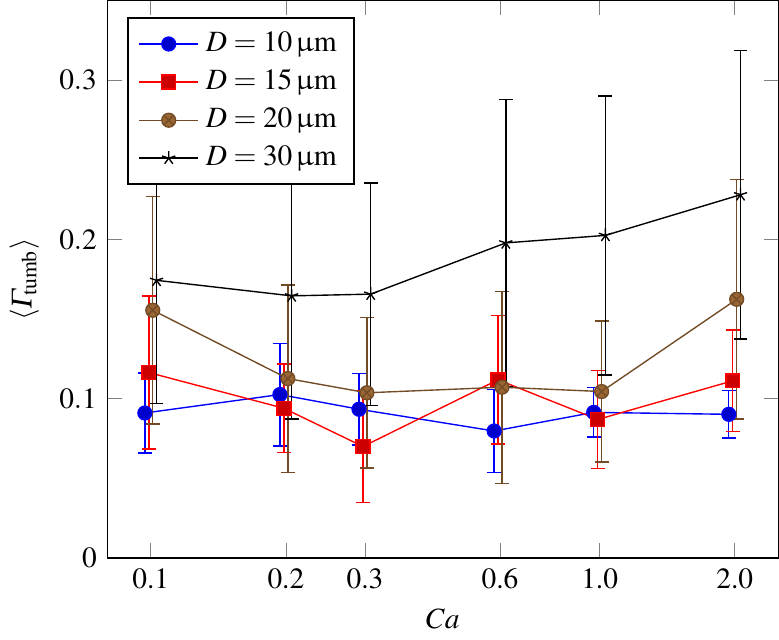}
 \caption{Normalised tumbling rate $\Gamma_\text{tumb}$ averaged over all platelets near the cell-free layer. Error bars indicate ensemble variances. Lines are guides for the eye. Note that the $\Canum$-axis is logarithmic and that symbols have been slightly shifted along the $\Canum$-axis to avoid overlap of error bars.}
 \label{fig:angularvelocity}
\end{figure}

Once platelets are caught in the vicinity of the CFL, they show two main dynamical states: i) tumbling in the local shear flow and ii) sliding parallel to the flow axis \citep{vahidkhah_platelet_2014}.
The central observable is the angular velocity of a platelet in a plane defined by the tube axis and the radial platelet position: $\omega_r$.
Fig.~\ref{fig:slidingtumbling} shows the time evolution of $\omega_r$ for two different platelets in the same simulation near the CFL.

A freely moving platelet with axis aspect ratio $p$ in a shear flow with shear rate $\dot{\gamma}$ would be in a pure tumbling state leading to Jeffery orbits with average angular velocity $\bar{\omega}_\text{J}$ \citep{jeffery_motion_1922}:
\begin{equation}
 \frac{\dot{\gamma}}{\bar \omega_\text{J}} = p + \frac{1}{p}.
\end{equation}
In the present case, $p = 3.6$, and therefore $\bar \omega_\text{J} = 0.26\, \dot{\gamma}$ as indicated by the dashed line in Fig.~\ref{fig:slidingtumbling}.
Sliding is caused by confinement due to the presence of the wall on one side of the platelet and the RBCs on the other.
Only occasionally a platelet has ``enough space'' to tumble.

As visible in Fig.~\ref{fig:slidingtumbling}, tumbling events are defined by localised peaks in $\omega_r$, while sliding is accompanied by low-amplitude fluctuations of $\omega_r$ with zero average.
Each tumbling event leads to a $180^\circ$- or $\pi$-rotation of the platelet, called a \emph{flip}.
In fact, the time integral of $\omega_r$ over one flip turns out to be close to $\pi$.
A full rotation would therefore require two flips.

An important parameter is the average relative tumbling rate of a platelet in the CFL:
\begin{equation}
 \label{eq:tumblingrate}
 \Gamma_\text{tumb} = \frac{\bar{\omega}_r}{\bar{\omega}_\text{J}} = \frac{1}{\bar{\omega}_\text{J}} \frac{\int f(t) \omega_r(t)\, \mathrm{d}t}{\int f(t)\, \mathrm{d}t}.
\end{equation}
To consider only platelets in proximity of the CFL, the function $f(t)$ is defined as
\begin{equation}
 f(t) = \begin{cases} 1 & r(t) > R - 2 \CFL \\ 0 & \text{otherwise}\end{cases}.
\end{equation}
Therefore, the integral in the denominator in Eq.~\eqref{eq:tumblingrate} is the total time a platelet spends near the CFL: $T_\text{CFL}$.

Fig.~\ref{fig:angularvelocity} shows the tumbling rate $\Gamma_\text{tumb}$ averaged over all platelets near the CFL.
The most striking observation is that tumbling is much less frequent than for a freely moving platelet.
The tumbling rate is rather reduced to $10$--$25\%$ of the value expected for a free platelet, indicating that sliding is the most likely dynamical state in the CFL for the investigated parameters.
\cite{vahidkhah_platelet_2014} reported that the sliding probability strongly increases with decreasing CFL thickness.
Below $\CFL \approx 4\,\micron$, which is the case here, sliding starts to occur until it eventually becomes the dominating dynamical mode.

From visual simulation data it follows that platelet flips are correlated with ``overtaking'' events of individual RBCs that move just at the edge of the CFL.
On the one hand, a platelet is dragged along by a passing RBC.
On the other hand, platelets find more space in the gaps between RBCs once an RBC has passed.
This makes it easier for a platelet to flip, but not all RBC overtaking events lead to a platelet flip.

Interestingly, $\Gamma_\text{tumb}$ is rather independent of $\Canum$ as seen in Fig.~\ref{fig:angularvelocity}.
The tube diameter has a stronger effect; larger tubes lead to a higher tumbling rate.
This is probably related to an increase of the CFL thickness with $D$: the thicker the CFL, the smaller the confinement felt by the platelets.
Flipping should therefore be easier and more frequent with increasing $D$ but otherwise constant flow parameters ($\Htt$, $\Canum$).

It is worth mentioning that the errors in Fig.~\ref{fig:angularvelocity}, defined by the variance of the ensemble average of all platelets, are relatively large.
The reason is that the total number of flips observed is small, and platelets show a wide distribution of number of flips.
On average, a platelet flips less than ten times during one simulation.

\section{Conclusions}
\label{sec:conclusions}

Tube flow of red blood cells (RBCs) and platelets was simulated using a combination of the lattice-Boltz\-mann, immersed boundary and finite element methods.
The aim of this work was to study platelet margination and subsequent platelet dynamics in the cell-free layer (CFL) for different tube diameters $D$ and RBC capillary numbers $\Canum$.
Tubes with diameters $D = 10$, $15$, $20$ and $30\,\micron$ were simulated, and the investigated capillary numbers range from $\Canum = 0.1$, where RBCs are relatively rigid, to $\Canum = 2.0$, where RBCs are strongly deformed and elongated.
The tube haematocrit is $\approx 37\%$.

One main result is that platelet margination is facilitated by a non-diffusive radial transport near the CFL edge when $\Canum > 0.2$.
This is in accordance with previous experimental observations.
For smaller $\Canum$, however, margination is less effective, and platelets are not irreversibly trapped in the CFL.

The non-diffusive platelet transport presents itself as a non-zero average platelet velocity in outward direction near the CFL edge.
Platelets within the outermost $5\,\micron$ in the tube are affected by this drift.
This explains why platelet margination is very fast for tubes with $D = 10$ and $15\,\micron$ where essentially all platelets are in wall vicinity.
Once a platelet has reached the CFL edge through diffusive motion, it is transported into the CFL due to collisions with tank-treading RBCs.
At least for $\Canum > 0.2$, this process is expected to continue until no platelets are left in the RBC-rich region.

The simulations also show that the radial platelet velocity fluctuations are strongly suppressed in the CFL.
The difference between these fluctuations within the RBC-rich region and the CFL increases with $\Canum$, which explains why platelets are less likely to escape the CFL with increasing $\Canum$.

Furthermore it is demonstrated that the predominant dynamical state of platelets in the CFL is sliding rather than tumbling, which is in line with earlier findings.
Due to the relatively small CFL thickness at $\approx 37\%$ haematocrit, platelets are strongly confined, and tumbling events are rare.
They occur $4$--$10$ times less frequently as expected for an unconfined platelet.
The tumbling rate is only a weak function of $\Canum$, but grows with $D$ and therefore the CFL thickness.

The presented results are in accordance with previous experiments and simulations.
As such, this work sheds more light on the platelet margination mechnism which is still not well understood.
It is demonstrated how the explicit modelling of deformable RBCs is necessary to observe margination as an emergent effect.
This research is thus expected to stimulate further discussions and contribute to a development of a predictive continuum margination theory.

\begin{acknowledgements}
 I acknowledge the award of a Chancellor's Fellowship from the University of Edinburgh and computer resources at Eindhoven University of Technology.
 There is no conflict of interest.
 Figures have been created with Ti\textit{k}Z and ParaView.
\end{acknowledgements}

% \bibliographystyle{spbasic}
% \bibliography{bibliography}

\end{document}